\documentclass[manuscript]{aastex}
\usepackage{epstopdf}

\shorttitle{SEARCH FOR GEV GRBs WITH ARGO-YBJ}
\shortauthors{BARTOLI ET AL.}

\begin{document}


\title{SEARCH FOR GEV GAMMA RAY BURSTS WITH THE ARGO-YBJ DETECTOR: SUMMARY OF EIGHT YEARS OF OBSERVATIONS}


\author{B.~Bartoli\altaffilmark{1,2},
 P.~Bernardini\altaffilmark{3,4},
 X.J.~Bi\altaffilmark{5},
 P.~Branchini\altaffilmark{6},
 A.~Budano\altaffilmark{6},
 P.~Camarri\altaffilmark{7,8},
 Z.~Cao\altaffilmark{5},
 R.~Cardarelli\altaffilmark{8},
 S.~Catalanotti\altaffilmark{1,2},
 S.Z.~Chen\altaffilmark{5},
 T.L.~Chen\altaffilmark{9},
 P.~Creti\altaffilmark{4},
 S.W.~Cui\altaffilmark{10},
 B.Z.~Dai\altaffilmark{11},
 A.~D'Amone\altaffilmark{3,4},
 Danzengluobu\altaffilmark{9},
 I.~De Mitri\altaffilmark{3,4},
 B.~D'Ettorre Piazzoli\altaffilmark{1,2},
 T.~Di Girolamo\altaffilmark{1,2,*},
 G.~Di Sciascio\altaffilmark{8},
 C.F.~Feng\altaffilmark{12},
 Zhaoyang Feng\altaffilmark{5},
 Zhenyong Feng\altaffilmark{13},
 Q.B.~Gou\altaffilmark{5},
 Y.Q.~Guo\altaffilmark{5},
 H.H.~He\altaffilmark{5},
 Haibing Hu\altaffilmark{9},
 Hongbo Hu\altaffilmark{5},
 M.~Iacovacci\altaffilmark{1,2},
 R.~Iuppa\altaffilmark{7,8},
 H.Y.~Jia\altaffilmark{13},
 Labaciren\altaffilmark{9},
 H.J.~Li\altaffilmark{9},
 G.~Liguori\altaffilmark{14,15},
 C.~Liu\altaffilmark{5},
 J.~Liu\altaffilmark{11},
 M.Y.~Liu\altaffilmark{9},
 H.~Lu\altaffilmark{5},
 L.L.~Ma\altaffilmark{5},
 X.H.~Ma\altaffilmark{5},
 G.~Mancarella\altaffilmark{3,4},
 S.M.~Mari\altaffilmark{6,16},
 G.~Marsella\altaffilmark{3,4},
 D.~Martello\altaffilmark{3,4},
 S.~Mastroianni\altaffilmark{2},
 P.~Montini\altaffilmark{6,16},
 C.C.~Ning\altaffilmark{9},
 M.~Panareo\altaffilmark{3,4},
 L.~Perrone\altaffilmark{3,4},
 P.~Pistilli\altaffilmark{6,16},
 F.~Ruggieri\altaffilmark{6},
 P.~Salvini\altaffilmark{15},
 R.~Santonico\altaffilmark{7,8},
 P.R.~Shen\altaffilmark{5},
 X.D.~Sheng\altaffilmark{5},
 F.~Shi\altaffilmark{5},
 A.~Surdo\altaffilmark{4},
 Y.H.~Tan\altaffilmark{5},
 P.~Vallania\altaffilmark{17,18,*},
 S.~Vernetto\altaffilmark{17,18},
 C.~Vigorito\altaffilmark{18,19,*},
 H.~Wang\altaffilmark{5},
 C.Y.~Wu\altaffilmark{5},
 H.R.~Wu\altaffilmark{5},
 L.~Xue\altaffilmark{12},
 Q.Y.~Yang\altaffilmark{11},
 X.C.~Yang\altaffilmark{11},
 Z.G.~Yao\altaffilmark{5},
 A.F.~Yuan\altaffilmark{9},
 M.~Zha\altaffilmark{5},
 H.M.~Zhang\altaffilmark{5},
 L.~Zhang\altaffilmark{11},
 X.Y.~Zhang\altaffilmark{12},
 Y.~Zhang\altaffilmark{5},
 J.~Zhao\altaffilmark{5},
 Zhaxiciren\altaffilmark{9},
 Zhaxisangzhu\altaffilmark{9},
 X.X.~Zhou\altaffilmark{13},
 F.R.~Zhu\altaffilmark{13},
 Q.Q.~Zhu\altaffilmark{5} and
 G.~Zizzi\altaffilmark{20}\\ (The ARGO-YBJ Collaboration)}


\altaffiltext{*}{Corresponding authors: digirola@na.infn.it, Piero.Vallania@to.infn.it, vigorito@to.infn.it}

 \altaffiltext{1}{Dipartimento di Fisica dell'Universit\`a di Napoli
                  ``Federico II'', Complesso Universitario di Monte
                  Sant'Angelo, via Cinthia, 80126 Napoli, Italy.}
 \altaffiltext{2}{Istituto Nazionale di Fisica Nucleare, Sezione di
                  Napoli, Complesso Universitario di Monte
                  Sant'Angelo, via Cinthia, 80126 Napoli, Italy.}
 \altaffiltext{3}{Dipartimento Matematica e Fisica "Ennio De Giorgi", 
                  Universit\`a del Salento,
                  via per Arnesano, 73100 Lecce, Italy.}
 \altaffiltext{4}{Istituto Nazionale di Fisica Nucleare, Sezione di
                  Lecce, via per Arnesano, 73100 Lecce, Italy.}
 \altaffiltext{5}{Key Laboratory of Particle Astrophysics, Institute
                  of High Energy Physics, Chinese Academy of Sciences,
                  P.O. Box 918, 100049 Beijing, P.R. China.}
 \altaffiltext{6}{Istituto Nazionale di Fisica Nucleare, Sezione di
                  Roma Tre, via della Vasca Navale 84, 00146 Roma, Italy.}
 \altaffiltext{7}{Dipartimento di Fisica dell'Universit\`a di Roma 
                  ``Tor Vergata'', via della Ricerca Scientifica 1, 
                  00133 Roma, Italy.}
 \altaffiltext{8}{Istituto Nazionale di Fisica Nucleare, Sezione di
                  Roma Tor Vergata, via della Ricerca Scientifica 1,
                  00133 Roma, Italy.}
 \altaffiltext{9}{Tibet University, 850000 Lhasa, Xizang, P.R. China.}
 \altaffiltext{10}{Hebei Normal University, Shijiazhuang 050016,
                   Hebei, P.R. China.}
 \altaffiltext{11}{Yunnan University, 2 North Cuihu Rd., 650091 Kunming,
                   Yunnan, P.R. China.}
 \altaffiltext{12}{Shandong University, 250100 Jinan, Shandong, P.R. China.}
 \altaffiltext{13}{Southwest Jiaotong University, 610031 Chengdu,
                   Sichuan, P.R. China.}
 \altaffiltext{14}{Dipartimento di Fisica dell'Universit\`a di 
                   Pavia, via Bassi 6, 27100 Pavia, Italy.}
 \altaffiltext{15}{Istituto Nazionale di Fisica Nucleare, Sezione di Pavia,
                   via Bassi 6, 27100 Pavia, Italy.}
 \altaffiltext{16}{Dipartimento di Fisica dell'Universit\`a ``Roma Tre'',
                   via della Vasca Navale 84, 00146 Roma, Italy.} 
 \altaffiltext{17}{Osservatorio Astrofisico di Torino dell'Istituto Nazionale
                   di Astrofisica, via P. Giuria 1, 10125 Torino, Italy.}
 \altaffiltext{18}{Istituto Nazionale di Fisica Nucleare,
                   Sezione di Torino, via P. Giuria 1, 10125 Torino, Italy.}
 \altaffiltext{19}{Dipartimento di Fisica dell'Universit\`a di 
                   Torino, via P. Giuria 1, 10125 Torino, Italy.}
 \altaffiltext{20}{Istituto Nazionale di Fisica Nucleare - CNAF, Viale
                   Berti-Pichat 6/2, 40127 Bologna, Italy.}


\begin{abstract}

The search for Gamma Ray Burst (GRB) emission in the energy range 1-100 GeV in coincidence with the satellite detection has been 
carried out using the Astrophysical Radiation with Ground-based Observatory at YangBaJing (ARGO-YBJ) experiment.
The high altitude location (4300 m a.s.l.), the large active surface ($\sim$ 6700 m$^2$ of Resistive Plate Chambers), the wide field of view ($\sim 2~$sr, limited only by the atmospheric absorption) and the high duty cycle ($>$ 86 \%) make the ARGO-YBJ experiment particularly suitable to detect short and unexpected events like GRBs.
With the scaler mode technique, i.e., counting all the particles hitting the detector with no measurement of the primary energy and arrival direction, the minimum threshold of $\sim$ 1 GeV can be reached, overlapping the direct measurements carried out by satellites.
During the experiment lifetime, from December 17, 2004 to February 7, 2013, a total of 206 GRBs occurring within the ARGO-YBJ field of view (zenith angle $\theta$ $\le$ 45$^{\circ}$) have been analyzed. This is the largest sample of GRBs investigated with a ground-based detector.
Two lightcurve models have been assumed and since
in both cases no significant excess has been found, the corresponding fluence upper limits in the 1-100 GeV energy region have been derived, with values as low as 10$^{-5}~$erg cm$^{-2}$. The analysis of a subset of 24 GRBs with known redshift has been used to constrain the fluence extrapolation to the GeV region together with possible cutoffs under different assumptions on the spectrum.

\end{abstract}


\keywords{gamma rays: bursts --- gamma rays: observations}



\section{INTRODUCTION}

Gamma Ray Bursts (GRBs) are among the most powerful sources in the sky, covering a very wide energy range from radio to multi-GeV $\gamma$-rays.
Even though they are located at cosmological distances
\citep{bib:costa97} at higher energies they outshine all other sources, including the Sun, during their typical duration of a  few seconds. 
GRBs occur at an average rate of a few per day, coming from the whole Universe.
Their high energy spectrum shows different features, the most important being a peak in the keV-MeV region.
There are at least two classes of GRBs, classified in terms of burst duration.
Short GRBs last up to 2 s and show a harder spectrum with a typical peak energy in the $\nu F_\nu$ spectrum at Earth at $\sim$ 490 keV \citep{bib:nava}.
Their origin is believed to be due to the merging of two compact objects like neutron stars or a neutron 
star and a black hole \citep{bib:ruffert, bib:rosswog}. 
 Recent support for this model comes from the optical and near-infrared detection of a faint transient, known as
``kilonova'', in the days following the short GRB130603B \citep{bib:tanvir13}.
Long GRBs have duration greater than 2 s with a softer spectrum and a typical $\nu F_\nu$ peak around 160 keV \citep{bib:nava}.
In this case the origin is believed to be due to the core collapse of type Ic supernovae, and indeed the coincidence
 of the two events has been observed in several cases (see for example \citet{bib:weiler01,bib:stanek, bib:gal04, bib:campana06}).
Most of the GRB spectra can be described by the Band function \citep{bib:band}, composed  of two smoothly joined power laws. This function fits quite successfully the convex shape and broad peak of the spectral energy distribution of the GRB prompt emission, however, being a phenomenological model, it does not take into account any physical explanation concerning  either the acceleration processes  or non-thermal radiative losses.
Despite the bulk emission is concentrated in the keV-MeV energy region,
EGRET \citep{bib:egret} and more recently Fermi \citep{bib:fermi} and AGILE \citep{bib:agile} satellites observed photons in the MeV-GeV range. 

At the time of writing this paper, the highest photon energy measured at 
Earth is 95 GeV, observed by the LAT instrument on the Fermi satellite from 
GRB130427A \citep{bib:zhu}.
The highest intrinsic energy ($\sim$ 147 GeV) detected from a GRB comes from a 27.4 GeV $\gamma$-ray observed during GRB080916C, 
which has a redshift of 4.35. This $\gamma$-ray was previously missed by the Fermi-LAT event analysis and was recently
recovered using an improved data analysis \citep{bib:atwood}.
Previously, the maximum observed photon energy was 33.4 GeV from 
GRB090902B ($\sim$ 94 GeV when corrected for its redshift $z$=1.822).
Up to now (May 2014) after almost 6 years of operation, Fermi-LAT 
detected photons above 10 GeV from one short (GRB090510) and 8 long GRBs
\citep{ bib:abdo09a, bib:abdo09b, bib:ackermann10, bib:ackermann11, bib:ackermann13, bib:zhu, bib:kocevski,
bib:vianello}.
Some of these GRBs (namely, GRB08916C, GRB090510, GRB090902B, GRB090926A, GRB130427A)
 cannot be well described  at GeV energies with an extrapolation of the
Band function seen at keV-MeV energies, but require a much harder energy 
spectrum starting from $\sim$ 100 MeV with a photon index $\alpha$ $\sim$ -2.

Another feature which characterizes the GeV emission is the light curve, with its onset delayed with respect to the
keV-MeV range and a longer duration, appearing as a very high energy afterglow.
The current models include emission in both 
internal \citep{bib:guetta03,bib:finke08} and external 
\citep{bib:kumar10,bib:ghisellini10,bib:ghirlanda10} 
shock scenarios, with $\gamma$-rays produced by leptonic or hadronic processes
via inverse Compton  scattering or neutral pion decay. 
The emission is believed to happen in highly relativistic narrow jets pointing towards the Earth.
The study of the GeV energy region could be of great help in discriminating between different models. 
As an example, the delayed onset of the high energy emission seen in most LAT-detected GRBs, if intrinsic, 
should favour the production from external shocks in the early GRB afterglow \citep{bib:fan} instead of the reverse shock
formed when the GRB ejecta encounter the interstellar medium \citep{bib:wang}.

GRBs have been detected through the whole universe, from the local one to redshift z=8.2, corresponding to
$\sim$ 95\% of the age of the universe.
Unfortunately, the energy resolution of the instruments onboard Fermi prevents the detection of clear spectral lines 
while their large angular uncertainty hampers the optical identification and follow-up.
For these reasons, only the GRBs seen in the keV-MeV region with arcmin resolution (as with
Swift-BAT) have a measured redshift.
In this same energy region the spectral index is usually measured but when the detected signal is weak also the
time-averaged spectrum is poorly constrained.
The absorption in the Extragalactic Background Light (EBL) greatly reduces the high energy photon flux from extragalactic sources.
The detection of $>$ 10 GeV photons from high redshift sources can be used to constrain the EBL amount from regions where it is highly uncertain.
Finally, the spectral slope in the GeV region could be of great help in discriminating between  
different GRB models. In particular, the detection of a cutoff energy could be indicative of e-pair production
at source allowing the measurent of the Lorentz boost factor of the jet \citep{bib:ackermann11}. On the other hand,
the spectral cutoff may be due to attenuation by the EBL, thus depending on the source redshift: GRBs at different
distances could be used to disentangle these two effects.

At present, all the experimental data in the MeV-GeV range have been obtained 
only from satellite detectors, which however, due to their limited size and the fast decrease of the source energy spectra, 
hardly cover the energy region above 1 GeV. 
Ground-based experiments can easily reach much larger effective areas exploiting two different techniques, which correspond to two different types of 
detectors: Extensive Air Shower (EAS) arrays and 
Imaging Atmospheric Cherenkov Telescopes (IACTs).
Concerning the latter, the huge telescope recently installed at the HESS site or the planned CTA
observatory can allow the detection of $\gamma$-rays with energy as low as 20-30 GeV \citep{bib:becherini, bib:konrad}, even if only at moderate zenith angles.
However, IACTs can operate only during nights with good weather conditions and no or limited
moon light, leading to a duty cycle of 10-15\%. Another disavantage is given by the limited full field of view, about $5^{\circ}$, which requires a fast 
slew after an external alert in order to observe a GRB, but
as pointed out by \citet{bib:gilmore}, the MAGIC experience shows that most observations started after considerably longer times despite the instrument rapid slew capabilities, with only a minority occurring with total delay times of $<$
$100~$s, preventing the detection of short GRBs and the study of the very prompt phase of long GRBs.
Due to the limited field of view, the prompt GRB location area must be quite small in order to be contained in it, but this is not the case for most of the GRBs detected by the Fermi-GBM.
Until now, all the major Cherenkov telescope arrays (MAGIC, HESS, VERITAS) attempted to detect a GRB with a
follow-up, but no robust positive result has been obtained and even with the new generation CTA only
$\lesssim$ 1 detection per year is expected \citep{bib:gilmore}.

On the contrary, EAS arrays have a large field of view ($\sim 2~$sr) and a very high duty cycle (in principle 100\%), however the requirement of a 
sufficient number of secondary particles in order
to reconstruct the shower arrival direction and primary energy leads to an energy threshold of at least $\sim$ 100 GeV. 
A possible technique to reduce the energy threshold of EAS detectors is working in scaler mode \citep{bib:vernetto} instead of shower mode, that is, recording 
the counting rates of the detector in search for an increase in
coincidence with a burst detected by a different experiment. Even if this technique does not
allow the reconstruction of the arrival direction and thus an 
independent search, it benefits from the large effective area and field of view and from the very low dead
time with an energy threshold typically around 1 GeV, thus overlapping the highest energies investigated by satellites experiments.
The resulting sensitivity is limited, but for GRBs observed at low zenith 
angles it is comparable to the highest fluxes measured by satellites.

The ARGO-YBJ detector has operated in scaler mode from December 17, 2004 to February 7, 2013.
In this period a total of 206 GRBs (selected from the GCN Circulars 
Archive\footnote{http://gcn.gsfc.nasa.gov/gcn3\_archive.html},
the Second Fermi GBM Gamma-Ray Burst Catalog \citep{bib:kienlin, bib:gruber} 
and the Fermi GBM Burst Catalog website\footnote{http://heasarc.gsfc.nasa.gov/W3Browse/fermi/fermigbrst.html}) in the field of view of the detector 
 were investigated searching for an increase in the detector counting rates. No significant excess has been 
found and corresponding upper limits to the fluence and energy cutoff under different assumptions on the spectrum are presented and discussed 
in this paper. A detailed 
description of the 
scaler mode technique, including the effective area calculation for gamma-rays and protons, the comparison between measured and simulated counting rates, the long-term counting rate behaviour and the detector stability over short and long time periods, can be found in \citet{bib:aielli08}, while the analysis procedure is described
in \citet{bib:aielli09}
together with the results on the first sample of GRBs analysed.
The GRB search can be done both in shower and scaler mode; here only
the results obtained with the latter are presented and discussed.
Shower mode results on a reduced sample of GRBs are given in \citet{bib:songzhan}.

\section{THE DETECTOR}


The Astrophysical Radiation with Ground-based Observatory at YangBaJing (ARGO-YBJ)
\citep{bib:aielli12}
is an EAS detector located at an altitude of
4300 m a.s.l. (atmospheric depth 606 g cm$^{-2}$) at the YangBaJing Cosmic 
Ray Laboratory (30.11$^{\circ}$N, 90.53$^{\circ}$E) in Tibet, P.R. China.
It is mainly devoted to $\gamma$-ray astronomy \citep{bib:aielli10, bib:bartoli11, bib:bartoli12a, bib:bartoli12b, bib:bartoli12c, bib:bartoli13a, bib:bartoli13b} and cosmic ray physics \citep{bib:aielli09c, bib:aielli11, bib:bartoli12d, bib:bartoli12e, bib:bartoli13c}.
The detector is made  of a single layer of Resistive Plate Chambers (RPCs), 
operated in streamer mode and grouped into 153 units named 
``clusters'', of size 5.7$\times$7.6 m$^2$ \citep{bib:aielli06}.
Each cluster is made by 12 RPCs (1.23$\times$2.85 m$^2$) and
each RPC is read out by 10 pads (55.6$\times$61.8 cm$^2$), 
representing the space-time pixels of the detector.
The clusters are disposed in a central full-coverage carpet (130 clusters on
an area 74 $\times$ 78$~$m$^2$ with $\sim$92\% of active surface) 
surrounded by a partially instrumented ($\sim$20\%) area up to 100 $\times$ 110$~$m$^2$, which 
increases the effective area and improves 
the reconstruction of the core location in shower mode.

In scaler mode the total counts are measured every $0.5~$s: for each cluster
the signal coming from its 120 pads
is added up and put in coincidence in a narrow time window (150 ns),
giving the counting rates for $\geq$ 1, $\geq$ 2, $\geq$ 3, and $\geq$ 4 
pads, which are read by four independent scaler channels.
These counting rates are referred in the following as 
$C_{\ge 1}$, $C_{\ge 2}$, $C_{\ge 3}$, and 
$C_{\ge 4}$, respectively, and 
the corresponding rates are $\sim$ 40 kHz, $\sim$ 2 kHz, $\sim$ 300
Hz, and $\sim$ 120 Hz. 
Since for the GRB search in scaler mode the authentication is only given
by a satellite detection, the stability of the detector and the
probability that it mimics a true signal are crucial and have to be deeply
investigated.

The main sources of counting rate variations are the pressure, acting 
on the shower 
development in the atmosphere, and the ambient temperature, acting on
 the detector efficiency.
The time scale of both variations is much larger than the typical 
GRB duration (seconds to minutes), so
they can be neglected provided that the behaviour of the single
 cluster counting rates is Poissonian.
A secondary local effect is due to the radon contamination in the detector 
hall. Electrons and $\gamma$-rays, from short-lived radon daughters 
(mainly $^{214}_{82}Pb,~^{214}_{83}Bi,~^{214}_{84}Po$) produced 
in the radon decay chain, are expected from $\beta$ decays and 
isotope de-excitations. It has been shown that they 
can influence the cluster counting rates at a level of a few per cent of the 
reference value. 
Even in this case the time variations are larger (hours) than the typical 
GRB duration and they can be neglected in the data processing 
\citep{bib:bartoli14, bib:radon}.

A very rapid variation can be induced by nearby lightning.
For this reason two electric field monitors {\it EFM-100}, located at 
opposite sides of the experimental hall, and a storm tracker {\it LD-250} 
(both devices by Boltek industries\footnote{http://www.boltek.com/}) 
have been installed to check the electric field variations.
Details of this study are widely discussed in \citet{bib:aef}.




\section{DATA SELECTION AND ANALYSIS}

\begin{figure}
\epsscale{.75}
\plotone{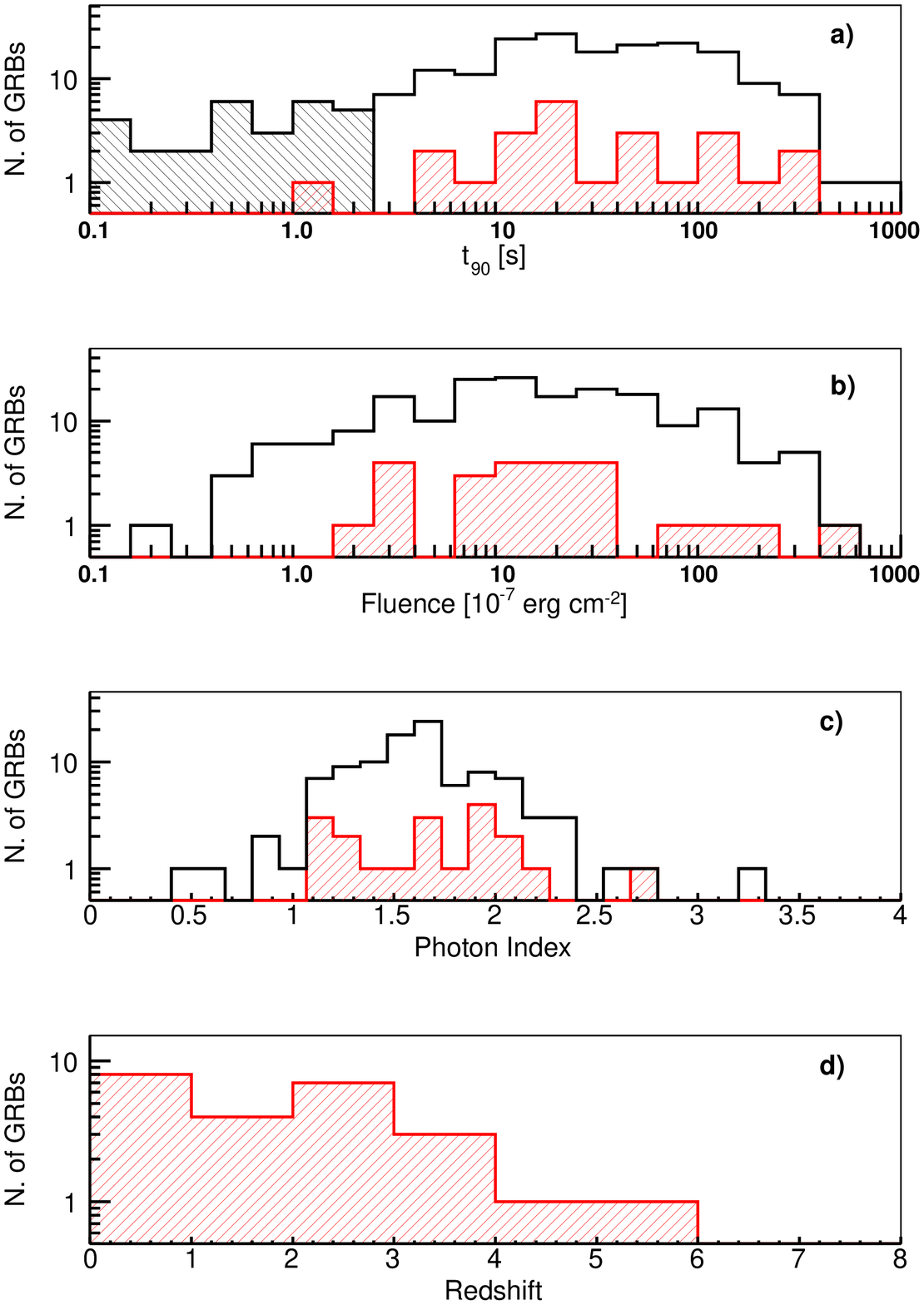}
\caption{Details of the GRB sample analyzed in coincidence with ARGO-YBJ:
a) $\Delta t_{90}$ durations of the whole sample (solid line) and of the GRBs 
with known redshift (filled area); 
b) Fluences measured by satellites (all normalized to the energy range 15-150 keV) for the whole available sample (solid line) and for the events with known redshift (filled area);
c) Photon index values  in the keV-MeV band for the whole available sample (solid line) and for the events with known redshift (filled area);
d) redshift values distribution. The dashed area on the left in plot a) 
indicates the short ($\leq$ 2s) GRB population.
\newline
(A color version of this figure is available in the online journal.)}
\label{fig:grbs}
\end{figure}

The ARGO-YBJ detector was completed in spring 2007, however, thanks to its 
modularity, the data taking started already in November 2004 (corresponding to
the launch of the Swift satellite), ending 
in February 2013, when the detector was definitively switched off. 
In this period a total of 223 GRBs, detected by satellite instruments, 
occurred inside the ARGO-YBJ field of view
(zenith angle $\theta$ $\leq 45^{\circ}$, corresponding to $1.84~$sr). 
The present analysis was carried out on 206 of them, since the other GRBs 
occurred during periods when the detector was inactive or not properly working.
Unlike $\Delta t_{90}$, defined as the time during which 90\% of the GRB 
keV-MeV photons is detected, the redshift and the spectrum in the same 
energy range are not always measured due to the difficulties 
introduced in Section 1.
The spectra measured by satellites can be fitted with a simple power law, a 
smooth double power law (Band or Smoothly Broken Power Law, 
SBPL \citep{bib:kaneko}) or a Cutoff Power Law (CPL). 
Figure \ref{fig:grbs}a shows the $\Delta t_{90}$ distribution, with the dashed 
area on the left indicating the short ($\leq$ 2s) GRB population,
while figure \ref{fig:grbs}b gives the distribution of the fluences measured
by the satellites, all normalized to the energy interval 15-150 keV. 
For 103 GRBs of our sample the simple power law
spectral index in the keV-MeV region was measured by satellite detectors 
and the corresponding distribution (with a mean value $<\alpha> =  1.6$) 
is shown in Fig. \ref{fig:grbs}c.
For 24 of them the redshift is also known and the corresponding distribution is shown in Fig. \ref{fig:grbs}d, being $<z> = 2.1$ the mean value of this subset. The durations $\Delta t_{90}$ and spectral indices $\alpha$ of GRBs with known 
redshift are pointed out in Fig. \ref{fig:grbs}a and \ref{fig:grbs} c, 
respectively, with a dashed area (coloured red in the online version). 
For this subset the mean value and width of the three distributions are 
compatible with those for the whole GRB sample.
The detailed list of the 24 GRBs with known redshift is given in 
table \ref{tab:subb}, while table \ref{tab:all} reports
the same information for the remaining 182 GRBs.

For each GRB the following standard procedure has been adopted: check of the 
detector stability, cluster selection by means of quality cuts and calculation 
of the significance of the coincident signal in the ARGO-YBJ detector.
In order to extract the maximum information from the experimental data, 
two analyses have been implemented:
\begin{itemize}
\item coincidence search for each GRB;
\item cumulative search for stacked GRBs.
\end{itemize}
Details on quality cuts
 and detector stability
are carefully discussed in \citet{bib:aielli08},
while the background evaluation and significance calculation, as well as the analysis technique itself, are described in 
\citet{bib:aielli09}.
\\

\subsection{\textit{Coincidence search}}
The counting rates of the clusters surviving the quality cuts 
(with an average efficiency over the whole data set $\sim 87 \%$)
are added up and the normalized fluctuation function
\begin{equation}
f = (c-b)/\sigma, ~~ \sigma = \sqrt{b+b\frac{\Delta t_{90}[s]}{600}}
\label{ffunction2}
\end{equation}
\\
is used to evaluate the significance of the excess observed in coincidence with the satellite detection,
where $c$ is the total number of counts in the $\Delta t_{90}$ time
window starting at $t_0$ (the trigger time) of the signal, both given by the 
satellite detector, and $b$ is the number of counts in a fixed time 
interval 
of $300~$s before and after the signal, normalized to the $\Delta t_{90}$ time.
This analysis can be done for the counting rates of all the multiplicities
$\ge$1, $\ge$2, $\ge$3, $\ge$4 and 1, 2, 3, where the counting rates $C_i$ 
are obtained from the measured counting rates $C_{\ge i}$ using the relation:

\begin{equation}
C_i = C_{\ge i} - C_{\ge i+1} \; \; \; (i=1,2,3)
\end{equation}

\begin{figure}
\plotone{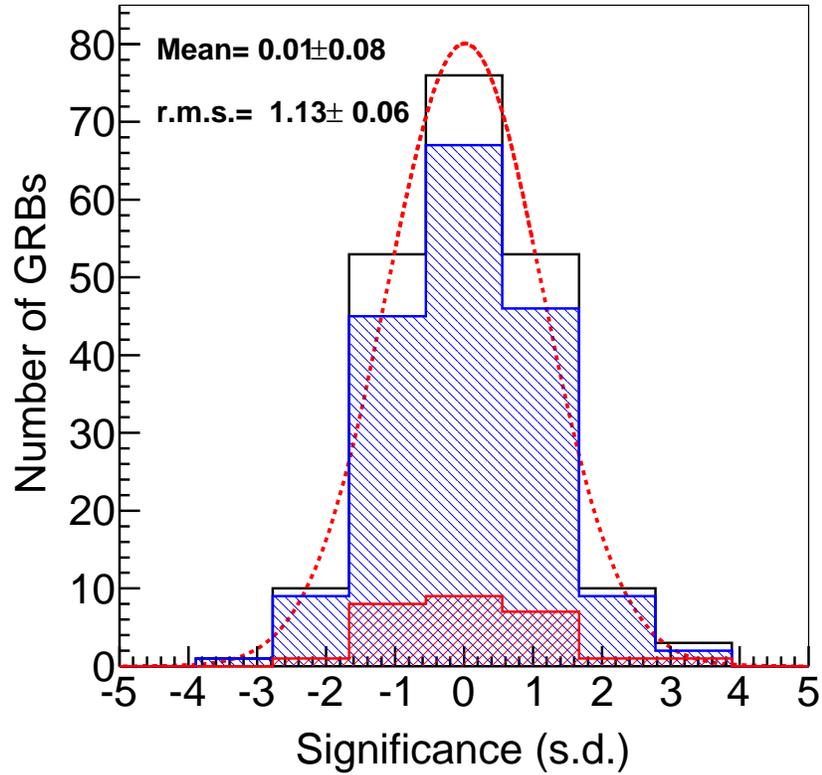}
\caption{Distribution of the statistical significances of the
206 GRBs with respect to background fluctuations (dark solid line) compared
with a free Gaussian fit (dotted line). Mean value and r.m.s. of the fit 
are shown. The light and dark dashed distributions refer to long and short 
GRBs, respectively.
\newline
(A color version of this figure is available in the online journal.)}
\label{fig:sigma}
\end{figure}

\noindent
In the following, if not otherwise specified, all the results are for the counting
rate $C_1$, which corresponds to the minimum primary energy in the
ARGO-YBJ scaler mode.
The detector stability over short time periods is discussed in \citet{bib:aielli08}, showing that the poissonian behaviour of the distribution of the normalized fluctuations $f$ is preserved provided that the total time window considered by the analysis (i.e., the signal interval $\Delta t_{90}$ plus the background interval 2$\times$300 s) is less than 30 minutes.
This condition is satisfied for all GRBs included in our data sample, therefore no long time corrections of the counting rates has been applied.
Even if the distributions of the single cluster counter rates for integrated times up to  half an hour are
Poissonian, this is not true for the sum of different clusters, which shows larger  fluctuations.
This effect has been carefully analysed and it was found due to the correlation between the counting rates of different 
clusters given by the air shower lateral distribution, i.e., counts in different clusters due to the same EAS  are
not independent. The resulting widening can be taken into account introducing
a Fano factor F \citep{bib:fano}:

\begin{equation}
\sigma^2 = F \sigma^2_p
\end{equation}

\noindent
where $\sigma^2_p$ is the Poissonian variance equal to the mean value of the counting rate distribution and 
$\sigma^2$ is the measured variance. The Fano factor increases with the number of detector units used and the integration time (i.e., the GRB duration)
while decreases for a sparse detector layout, and its effect is to reduce the sensitivity by a factor
$\sqrt F$. For each GRB the $\sqrt F$ is listed in tables \ref{tab:subb} and \ref{tab:all}, and the mean value calculated over the whole data sample is
 $<\sqrt F> = 2.22$.
In order to take into account this effect and calculate properly the signal significance we studied the local fluctuation of the  normalized function $f$ (defined in equation \ref{ffunction2})
in an interval $\pm12~h$ around the GRB trigger time and used equation (17) of  \citet{bib:li_ma83}.
Figure \ref{fig:sigma} (dark solid line) shows the distribution of the resulting significances for all the 206 GRBs.
No significant excess is measured, the largest being 3.52$\sigma$ for
GRB080727C, with a
post-trial chance probability of $4.5 \cdot 10^{-2}$. Since the long GRBs typically show a softer spectrum with a lower Band peak energy, the same distribution only for the 27 short GRBs is shown in the same figure \ref{fig:sigma} (dark dashed area, coloured red in the online version).
Even in this case no significant excess is measured, the most significant event being GRB051114 with
3.37$\sigma$ and a post-trial chance probability of $1.0 \cdot 10^{-2}$.
For this GRB, since we expect a harder energy spectrum from short GRBs, we carried out the same analysis using the higher multiplicity channels $C_2, C_3$ and $C_{\ge 4}$, obtaining a significance of 1.16, 1.09 and 1.95$\sigma$, 
respectively.

\begin{figure}
\plotone{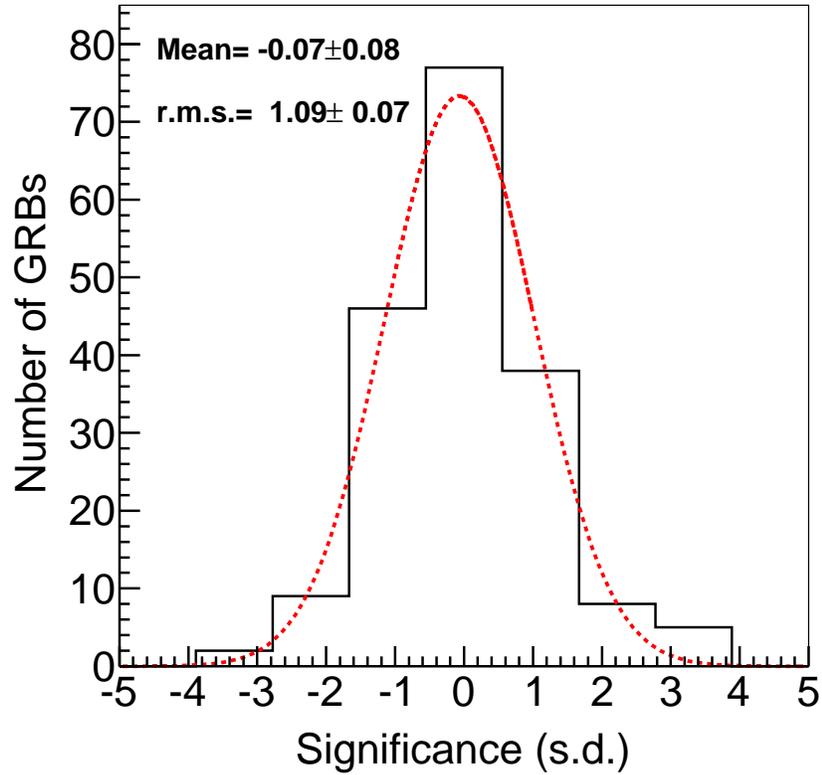}
\caption{Distribution of the statistical significances of the
185 GRBs with $\Delta t_{90} \ge 1.5~$s with respect to background fluctuations (solid line) 
compared with a free Gaussian fit (dotted line) for the extended time 
window search (see text). Mean value and r.m.s. of the fit are shown.
\newline
(A color version of this figure is available in the online journal.)}
\label{fig:sigma_ext}
\end{figure}

Besides this search, a time window broader than $\Delta t_{90}$ has been considered to take into account the possible high energy afterglow.
\citet{bib:ghisellini10} found that the flux of 
8 among the 11
brightest bursts detected by Fermi-LAT above 100 MeV (in the first 13 months of operation) decays as a power law with a typical slope $t^{-1.5}$.
In this analysis we assumed this trend in the afterglow phase ($t \geq \Delta t_{90}$) and a constant flux during the GRB prompt emission since we consider only the time-averaged behaviour:
%
%
%
\begin{eqnarray}
&A(t)=A_0 \; \; \; &(t \leq \Delta t_{90})\\
\nonumber
& \; \; \; \; \; \;\; \; \; \; \; \; \; \; \;\; \; \; A(t)=A_0(t/\Delta t_{90})^{-3/2} \; \; \; &(t > \Delta t_{90})
\end{eqnarray}
with $A_0$ corresponding to the mean flux during the low energy emission time $\Delta t_{90}$.
With this assumption, 2/3 of the total emission comes after $\Delta t_{90}$.
To search for such a delayed emission, a longer time interval $\Delta t'_{90}$ has to be used.
Its value is chosen in order to maximize the signal significance.
Assuming Poissonian fluctuations and introducing a mean background counting rate $k$ in units of $\Delta t_{90}$, the significance is:
\begin{equation}
\sigma(t)=\frac{\int_0^t A(t)dt}{\sqrt{k \cdot t/\Delta t_{90}}}
\end{equation}
The maximum of this function is at $t/\Delta t_{90}=16/9$.
In our case, since the fluctuations are not purely Poissonian and the Fano factor F depends on the integration time, we searched for a maximum significance of the modified function:
\begin{equation}
\sigma'(t)=\sigma(t)/\sqrt{F(t)}
\end{equation}
with an iterative procedure, increasing for each GRB the time window by the minimum 0.5$~s$ step. Then the Fano factor is calculated giving the resulting significance from equation (6).
This procedure is repeated covering the time interval from $\Delta t_{90}$ to $2 \Delta t_{90}$.
The significance curve in this time window is then fitted by a second-order polynomial and the $\Delta t'_{90}$ corresponding to its maximum is used instead of $\Delta t_{90}$ for this extended search.
Since the Fano factor increases with time, $\Delta t'_{90}$ is always shorter than the purely Poissonian value and certainly fall into the search interval.
This procedure searches for a maximum in the [$\Delta t_{90}$ -- $2 \Delta t_{90}$] range in steps of 0.5$~$s, therefore 
the analysis has been limited to GRBs 
with $\Delta t_{90} \ge 1.5~$s, allowing a second order fit of function (6).
Moreover, for the longer GRBs the Fano factor is so big that the increase of $\Delta t_{90}$ does not improve the sensitivity 
and a clear maximum cannot be found.
For these events $\Delta t'_{90}=\Delta t_{90}$ has been used, since this is the value that maximizes the signal to noise ratio for a constant signal during $\Delta t_{90}$.
The $\Delta t'_{90}$ obtained for the 185 GRBs with $\Delta t_{90} \ge 1.5~$s are listed in tables \ref{tab:subb} and \ref{tab:all} (for 61 of them $\Delta t'_{90}=\Delta t_{90}$).
The corresponding significance distribution is shown in figure 
\ref{fig:sigma_ext}. 
No significant excess is found also in this case, the larger one being 
3.52$\sigma$ for GRB080727C, with a post-trial chance probability of $4.1 \cdot 10^{-2}$.

\subsection{\textit{Stacked analysis}}
Besides the coincidence analysis for each GRB, a stacked analysis has been 
carried out in order to search for common features of all GRBs in 
{\it Time} or in {\it Phase}.

In the {\it Time} analysis the counting rates for all the GRBs, 
in 9 windows ($\Delta t=0.5,1,2,5,10,20,50,100$ and $200~$s) starting at $t_0$,
have been added up in order to investigate 
a possible common duration of the high energy emission. 
A positive observation at a fixed $\Delta t$ could be used as an alternative 
value to the observed $\Delta t_{90}$ duration and a different way to look 
for a possible high energy delayed component.
Since the bins are not independent, the distribution of the significances of the 9 time intervals is compared with random distributions obtained for starting times different from $t_0$ in a time interval $\pm$12 hr around the true GRB 
trigger time.
Moreover, for the sample of GRBs with known redshift (with z ranging from 0.48 to 5.6) the time windows have been corrected for the cosmological dilation factor $(1+z)$.
The most significant excess ($1.5\sigma$) is observed for the sample of 182 GRBs with no redshift at 
$\Delta t=0.5~$s with a chance probability of 0.60, while the analysis of the 24 GRBs with measured redshift led to a maximum significance of $0.7\sigma$ in the shorter time window ($\Delta t=0.5~$s at z=0).

In the {\it Phase} analysis only 165 GRBs with duration $\Delta t_{90}\geq 5~$s
have been added up scaling their duration to a common phase plot 
(i.e., 10 bins each sampling a $10\%$ wide interval of $\Delta t_{90}$, 
being $0.5~$s the minimum duration for the scaler mode data acquisition). 
This analysis should point out a common feature of all GRBs in case of a GeV 
emission correlated with the GRB duration at lower energy.
Even in this case no excess is found, and the most significant 
bin, corresponding to the phase $[0.7-0.8]$ of $\Delta t_{90}$, 
has a marginal significance of 1.78$\sigma$.

%






\section{FLUENCE AND E$_{CUT}$ UPPER LIMITS}

\begin{figure}
\plotone{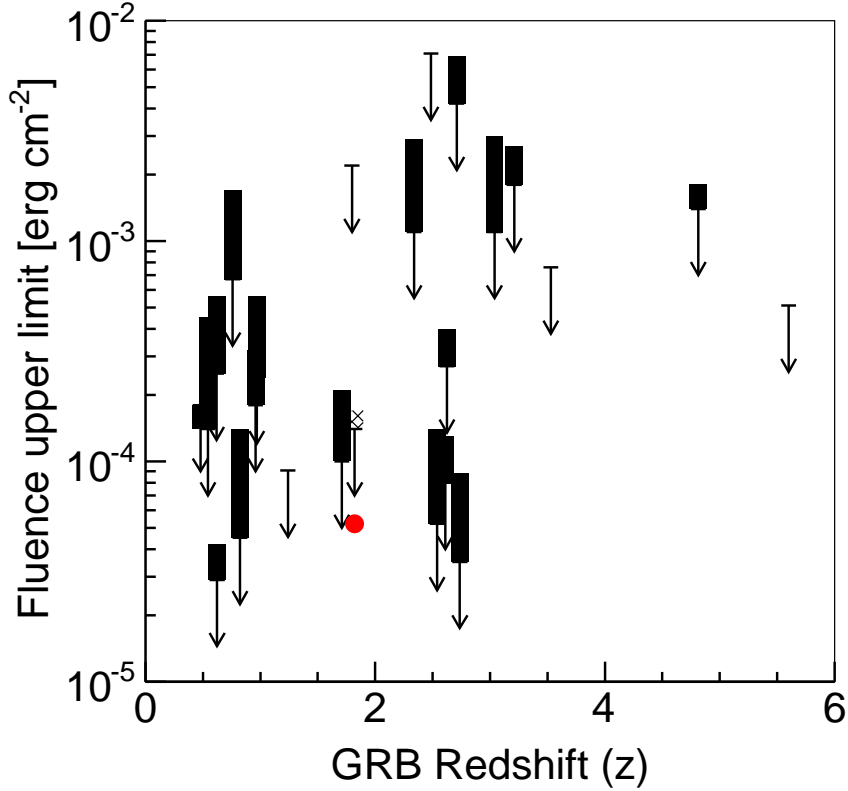}
\caption{Fluence upper limits of GRBs in the 1-100 GeV interval as a function 
of redshift. The rectangles represent the values obtained with
differential spectral indexes ranging from $\alpha=-2.5$ to the
satellite measurement $\alpha_{sat}$. The 5 arrows give the upper limits
for the former case only, these GRBs being best fitted at lower energies 
with a cutoff power law spectrum. The dot shows the fluence extrapolated 
in the 1-100 GeV range from the Fermi-LAT observations of GRB090902B; 
only for this GRB the GeV spectral index has been used and the dashed area 
has been obtained applying an energy cutoff running from 30 to 100 GeV.
\newline
(A color version of this figure is available in the online journal.)}
\label{fig:ul1}
\end{figure}

The fluence upper limits can be derived in the $[1-100]~GeV$ range 
from our experimental data 
and making some assumptions on the GRB primary spectrum.
For this calculation we used the maximum number of counts at 99\% 
confidence level (c.l.) following equation (6) of \citet{bib:helene}.
The interaction of the GRB photons with the EBL results in e-pair production 
which originates a spectral cutoff.
This effect depends on the GRB redshift, with a lower cutoff energy for more 
distant GRBs.
For this reason, the most meaningful upper limits are obtained for the
sample of 24 GRBs with known redshift (see table \ref{tab:subb}), while 
for the others (table \ref{tab:all}) a value of z=2 and z=0.6 has been adopted for long and short GRBs, respectively,
according to their measured distributions \citep{bib:jakobsson06, bib:berger05, bib:berger13}.
For the differential spectral indexes we used two extrapolations to estimate 
the expected high energy fluence for each GRB: 
a) the spectral index $\alpha_{sat}$ measured 
by satellite detectors in the keV-MeV energy range 
(corresponding to the $f_{sat}$ values in tables 
\ref{tab:subb} and \ref{tab:all}) and 
b) the conservative value $\alpha=-2.5$ 
($f_{2.5}$ values in tables \ref{tab:subb} and \ref{tab:all}). 
For case a), when the Band or SBPL spectral features have been identified, 
the higher energy spectral index (i.e. above the peak in the keV-MeV region) 
has been used. These assumptions represent respectively the most and less 
favourable spectral index hypotheses.
The absorption effect due to the EBL is taken into account using the model 
described in \citet{bib:kneiske04} and
applying an exponential cutoff to the spectrum according to the redshift. 
Figure \ref{fig:ul1} shows the $99\%~$c.l. upper limits as a function of z 
for the GRBs with known redshift.
For 5 of them, whose spectrum is best fitted by a CPL, 
only the upper limits for case b) are given.

For GRB090902B (which was the GRB in the ARGO-YBJ field of view 
with the highest energy photon detected) the fluence extrapolated from 
Fermi-LAT observations in the same energy range is shown. 
Only for this GRB the GeV spectral index measured by Fermi-LAT has been used 
and the dashed area in
figure \ref{fig:ul1} has been obtained applying an energy cutoff to the GRB 
spectrum running from 30 GeV (about the maximum energy measured by Fermi-LAT) 
to 100 GeV.
According to our calculation, in the case of a spectrum extending up to 
100 GeV the extrapolated GRB fluence 
is just a factor 2.7 lower than our expected sensitivity.
Due to the peculiar GeV emission of this GRB, the search has been done also in different time windows, in particular in coincidence 
with the extended  Fermi-LAT emission [$0-90~$s], the maximum density of events with energy $>$1 GeV [$6-26~$s] and the time of the 33.4 GeV photon [$82-83~$s]. 
The resulting significances are -0.03, 1.00 and -0.52$\sigma$, respectively.

\begin{figure}
\plotone{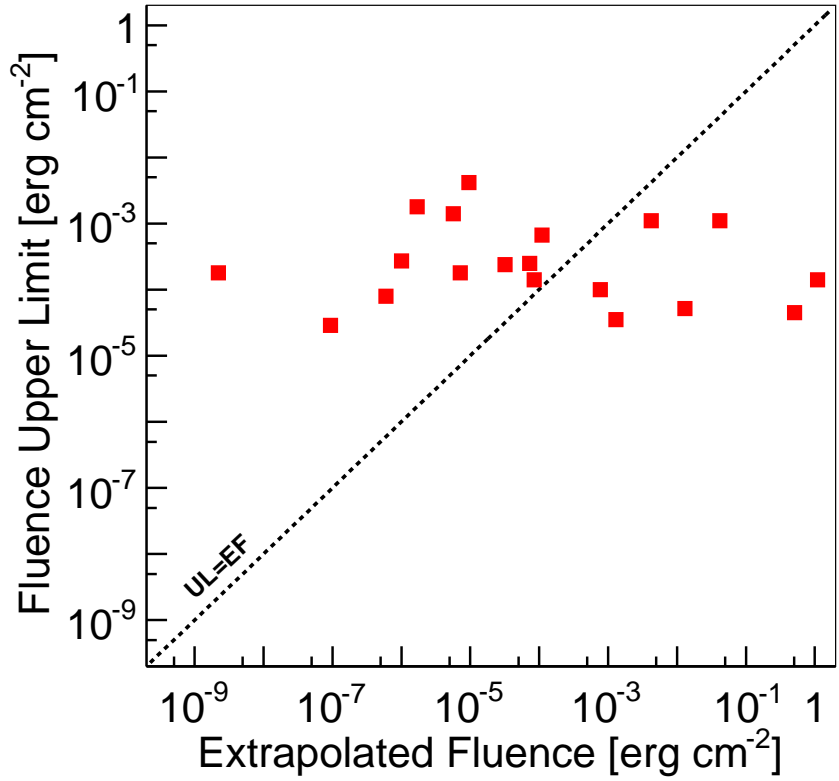}
\caption{ARGO-YBJ upper limits (in the 1-100 GeV interval) vs. fluence 
extrapolation for GRBs with measured redshift and low energy power-law
spectral index.
\newline
(A color version of this figure is available in the online journal.)}
\label{fig:ul2} 
\end{figure}

\begin{figure}
\plotone{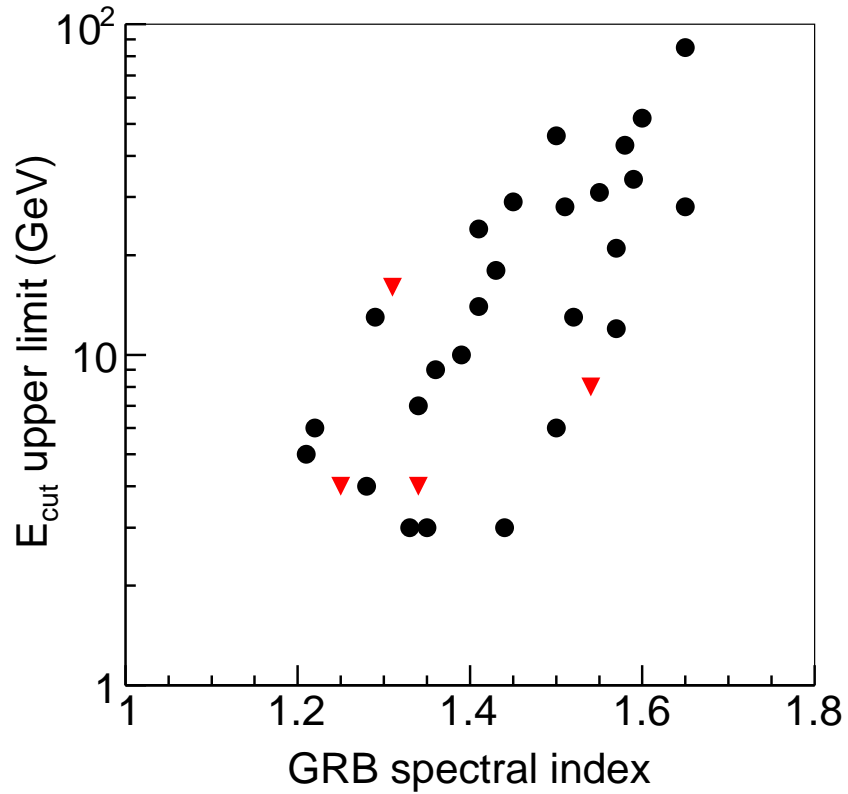}
\caption{Cutoff energy upper limits as a function of the spectral index obtained by extrapolating the measured keV spectra.
The values represented by the triangles are obtained taking into account extragalactic absorption
 at the known GRB redshift. For the other GRBs (dots) z=2 and z=0.6 are assumed for
long e short ones, respectively.
\newline
(A color version of this figure is available in the online journal.)}
\label{fig:ecut_ul} 
\end{figure}

A comparison between the expected fluence, obtained extrapolating the
keV-MeV spectra measured by satellites and including the EBL absorbtion, 
and the fluence upper limit determined with the ARGO-YBJ scaler data 
has been done for the 19 GRBs with measured redshift and energy spectrum 
best fitted by a simple power law, excluding the 5 events which present a  
CPL spectrum. The result is shown in figure \ref{fig:ul2}.
The 7 points on the right side of the line 
{\it Upper Limit (UL)= Expected Fluence (EF)} 
(i.e. in the region where the upper limits are lower than the 
extrapolated fluences) indicate that, 
since the corresponding GRBs were not detected, 
the chosen extrapolation is not feasible up to our range [$1-100~$GeV] 
or a cutoff should be present in the high energy tail of the spectrum.
Therefore, assuming the spectral index measured at low energies, the 
maximum cutoff energy has been estimated as follows.
The extrapolated fluence is calculated together with the fluence upper 
limit as a function of the cutoff energy $E_{cut}$. 
If the two curves cross in the [$2-100~$GeV] interval, 
the intersection gives the upper limit to the cutoff energy. 
This is what happens to four of them (GRB050802, GRB081028A, GRB090809A and GRB110128A), for which the knowledge of the redshift allows the estimation of the extragalactic absorption and hence a more accurate fluence upper limit and cutoff energy determination. 
For three of them (GRB071112C, GRB090424 and GRB130113B) the estimated $E_{cut}$ upper limit is below $2~$GeV:
 we can conclude that in these cases the low energy spectrum cannot be 
extended to the GeV region and some additional features occur in the 
keV-MeV range.
The values obtained for $E_{cut}$ are reported in the last column of 
table \ref{tab:subb} and shown in figure \ref{fig:ecut_ul} (triangles)
as a function of the spectral index.
The same calculation  can be made for the GRBs with unknown redshift assuming for the EBL absorbtion
z=2 and z=0.6 for long and short ones respectively and the resulting $E_{cut}$ values are given in 
the last column of table \ref{tab:all}
and shown in figure \ref{fig:ecut_ul} (dots).
More realistic models for the spectrum shape and/or different hypotheses on the 
photon spectral index in the GeV region can be considered.
Since all the 7 GRBs falling on the right side of the {\it UL=EF} line in 
Fig. \ref{fig:ul2} are long, we first assumed a Band spectrum with an 
$E_{peak}$ value of 160 keV and a spectral index
$\beta=-2.34$, corresponding to the mean peak energy and high energy slope 
for this class of GRBs \citep{bib:nava}.
With this model all the 7 GRBs result under threshold (i.e., the extrapolated 
fluence is lower than our upper limit).

Another possibility is to suppose a fixed ratio between the GeV and 
keV-MeV fluences. The simultaneous observation
of GRBs in these energy bands has been performed in the past by EGRET and 
BATSE onboard the CGRO satellite and more recently 
by Fermi-LAT and Fermi-GBM for a handful of events.
As pointed out by \citet{bib:dermer10}, for long GRBs this ratio is close to 0.1 when 
the energy ranges considered to determine the fluence are 100 MeV - 10 GeV 
and 20 keV - 2 MeV. As the GeV spectral index we used a value -2,
consistent with both EGRET and  Fermi-LAT mean values.
This high energy component represents a strong deviation with respect to the 
Band spectrum, increasing significantly the expected GeV fluence
 even if to a smaller extent than extrapolating the keV-MeV spectra.
Also under these hypoteses all the 7 long GRBs fall on the left side of the 
{\it UL=EF} line in Figure \ref{fig:ul2}.

\section{DISCUSSION AND CONCLUSIONS}

The detection of high energy photons by the Fermi-LAT instrument clearly 
demonstrates that at least a small fraction of GRBs emits in the GeV range. 
The detected photons experience two main processes: generation at the 
source and propagation through the intergalactic medium. 
Several models have been proposed to explain the production of high energy 
photons in GRBs, but according to the standard fireball shock model they are
essentially caused by internal or external shocks.
Once produced, a fraction of these photons are converted into electron-positron pairs due 
to the interaction with low energy photons, mainly of the 
infrared-optical-ultraviolet cosmic background (EBL).
This mechanism limits the photon mean free path and thus the visible horizon, 
which decreases with the energy up to $\sim$ 10$^{15}$ eV, where
the interaction with the Cosmic Microwave Background Radiation makes
 it smaller 
than the Galactic radius.
The signal reaching the Earth is the final result of all these production and 
propagation mechanisms, bringing valuable information on all of them 
but at the same time difficult to separate.
Features like the maximum energy as a function of the redshift, the photon 
index and other temporal and spectral characteristics, if seen with sufficient 
statistics, could discriminate between different mechanisms and shed 
light in this still largely unknown field.
For these reasons the study of GRBs would greatly benefit from the 
contribution of ground-based detectors to the direct satellite measurements.

In this paper a search for GRBs in coincidence with satellite detections has 
been carried out using the complete ARGO-YBJ data set.
During about 8 years a total of 206 GRBs has been analysed, producing the 
largest GRB sample ever studied using the scaler mode technique.
In the search for GeV $\gamma$-rays in coincidence with the GRB satellite 
detections, no evidence of emission was found for any event both for the 
whole sample and for separate analyses of the two populations of long and 
short GRBs.
For GRBs with duration $\ge 1.5~$s the search for a signal in a time window extended with respect 
to the low energy one has been carried out with similar results.
The stacked search, both in time and phase, has shown no deviation from the 
statistical expectations.
The subset of 24 GRBs with known redshift has been carefully analysed in 
terms of fluence and cutoff upper limits.
For GRB090902B the fluence upper limit using the GeV spectral index is very 
close to the  Fermi-LAT measurement (a factor 2.7 higher), supposing a 
high energy emission extending from the observed 30 GeV up to 100 GeV.
This GRB was certainly our best candidate for a detection, however an area 
7.2 times larger would have been necessary.
For the other GRBs with known redshift, fluence upper limits as low as 
$2.9 \times 10 ^{-5}$ erg cm$^{-2}$ in the 1-100 GeV energy range have been 
set, assuming an high energy spectral index equal to that measured by
satellites. Under this hypothesis, for 7 of them an upper limit to the cutoff 
energy has also been determined, otherwise an average Band spectrum or a fixed 
ratio between the high and low energy fluences must be assumed.

The expected rate of GRBs which could be observed by the ARGO-YBJ experiment, 
based on the Swift satellite detections, was between 0.1 and 0.5 year$^{-1}$ 
\citep{bib:aielli08} and it should have doubled with the later launch of the 
Fermi satellite. 
The value of $0.3~$year$^{-1}$ obtained for our 90\% c.l. upper limit is close 
to our lower expectation partially because the predicted Fermi 
detection rate was overestimated and partially because the LAT-detected GRBs 
have a spectrum softer than presumed.

In the next future, three huge ground-based detectors could continue this 
search with improved sensitivity. HAWC, a water Cherenkov detector with a 
surface of 22000 m$^2$ is under construction in Mexico at an altitude of 
4100 m a.s.l.. Its expected detection rate is 1.55 year$^{-1}$ for short 
GRBs and 0.25 year$^{-1}$ for long GRBs, mainly using the shower 
mode technique in the range 50-500 GeV \citep{bib:taboada}.
CTA will observe the night sky detecting the atmospheric Cherenkov light.
Its huge telescopes for the detection of low energy $\gamma$-rays have been 
designed also for a fast slewing, allowing a repointing time $\lesssim 100~$s.
Apart from a very lucky serendipitous observation, the CTA search is limited 
to long GRBs after the very prompt phase, with an expected detection rate 
ranging from 0.6 to 2 year$^{-1}$ according to baseline or optimistic 
assumptions and with a strong dependence on the energy threshold 
(more than on the pointing delay) \citep{bib:gilmore}.
GRB detection from ground via the water Cherenkov technique will also be 
possible with the proposed LHAASO experiment \citep{bib:lhaaso}, 
whose detection rate has not yet been estimated.
Thirty years after the first proposal by \citet{bib:morello},
the first solid detection of a GRB from ground seems at hand.




\acknowledgments

This work is supported in China by NSFC (Grant
No. 10120130794), the Chinese Ministry of Science and
Technology, the Chinese Academy of Sciences, the Key
Laboratory of Particle Astrophysics, CAS, and in Italy by
the Istituto Nazionale di Fisica Nucleare (INFN). We also
acknowledge the essential support of W.Y. Chen, G. Yang,
X.F. Yuan, C.Y. Zhao, R. Assiro, B. Biondo, S. Bricola,
F. Budano, A. Corvaglia, B. D’Aquino, R. Esposito,
A. Innocente, A. Mangano, E. Pastori, C. Pinto, E. Reali,
F. Taurino, and A. Zerbini in the installation, debugging,
and maintenance of the detector.

\clearpage



\clearpage

\clearpage

\clearpage

\clearpage

\clearpage

\clearpage

\begin{deluxetable}{clcccccccrrccc}
\tabletypesize{\scriptsize}
\rotate
\tablecaption{GRBs with measured redshift observed by ARGO-YBJ.
}
\tablewidth{0pt}
\tablehead{
\colhead{GRB} & \colhead{Satellite} & 
\colhead{$\Delta t_{90}$} &
\colhead{$\Delta t'_{90}$} &
\colhead{$\theta $} &  
\colhead{$z$} &
\colhead{$\sqrt{F}$} &
\colhead{Spectral Index} & 
\colhead{$A_{det}$} & \colhead{$\sigma$} & \colhead{$\sigma'$} &
\colhead{Fluence U.L.\tablenotemark{a,c}} &
\colhead{Fluence U.L.\tablenotemark{b,c}} &
\colhead{$E_{cut}$ U.L.\tablenotemark{d}}\\
\colhead{} & \colhead{} & \colhead{(s)} & \colhead{(s)} & \colhead{($^{\circ}$)}  & \colhead{}  &  \colhead{} & \colhead{($\alpha_{sat}$)} & \colhead{(m$^2$)} & \colhead{}  & \colhead{}  & \colhead{(erg cm$^{-2}$) ($f_{sat}$)} & \colhead{(erg cm$^{-2}$) ($f_{2.5}$)} & \colhead{(GeV)}
\\
\colhead{(1)} & \colhead{(2)} & \colhead{(3)} &
\colhead{(4)} & \colhead{(5)} & 
\colhead{(6)} & \colhead{(7)}  & \colhead{(8)} & \colhead{(9)} &
\colhead{(10)} & \colhead{(11)} & \colhead{(12)} &
\colhead{(13)} & \colhead{(14)} 
}

\startdata
050408 & HETE & 15 &  23.7 & 20.4 & 1.24 & 1.3 & CPL & 1560 & -2.12 & -2.90 &$  - $&$ 9.1\cdot 10^{-5} $&  - \\
050802 & Swift & 19 &  31.4 & 22.5 & 1.71 & 1.2 & 1.54 & 1516 & 0.19 & -0.02 &$ 1.0\cdot 10^{-4} $&$ 2.1\cdot 10^{-4} $& 8 \\
060115 & Swift & 139.6 & 139.6 & 16.6 & 3.53 & 3.0 & CPL & 3985 & -1.02 & -1.02 &$  - $&$ 7.6\cdot 10^{-4} $&  - \\
060526 & Swift & 298 & 298.0 & 31.7 & 3.21 & 3.9 & 2.01 & 4029 & -1 & -1.00 &$ 1.8\cdot 10^{-3} $&$ 2.7\cdot 10^{-3} $&  - \\
060714 & Swift & 115 & 115.0 & 42.8 & 2.71 & 4.9 & 1.93 & 5155 & -0.61 & -0.61 &$ 4.2\cdot 10^{-3} $&$ 6.9\cdot 10^{-3} $&  - \\
060927 & Swift & 22.5 &  31.3 & 31.6 & 5.6 & 1.6 & CPL & 5242 & -0.14 & -0.47 &$  - $&$ 5.1\cdot 10^{-4} $&  - \\
061110A & Swift & 40.7 &  40.7 & 37.3 & 0.76 & 2.5 & 1.67 & 5545 & 0.01 &  0.01 &$ 6.7\cdot 10^{-4} $&$ 1.7\cdot 10^{-3} $&  - \\
071112C & Swift & 15 &  22.0 & 18.4 & 0.82 & 1.5 & 1.09 & 5198 & 1.01 &  0.46 &$ 4.5\cdot 10^{-5} $&$ 1.4\cdot 10^{-4} $& $<2$ \\
081028A & Swift & 260 & 260.0 & 29.9 & 3.04 & 3.9 & 1.25 & 5805 & 0.37 &  0.37 &$ 1.1\cdot 10^{-3} $&$ 3.0\cdot 10^{-3} $& 4 \\
090424 & Swift & 48 &  57.2 & 33.1 & 0.54 & 2.0 & 1.19 & 5762 & 0.6 &  0.71 &$ 1.4\cdot 10^{-4} $&$ 4.5\cdot 10^{-4} $& $<2$ \\
090426 & Swift & 1.2 &  - & 43.7 & 2.61 & 1.3 & 1.93 & 5805 & -1.08 &  - &$ 8.0\cdot 10^{-5} $&$ 1.3\cdot 10^{-4} $&  - \\
090529A & Swift & 100 & 107.1 & 19.9 & 2.63 & 2.2 & 2 & 5892 & -0.66 & -0.83 &$ 2.7\cdot 10^{-4} $&$ 4.0\cdot 10^{-4} $&  - \\
090809A & Swift & 5.4 &  7.9 & 34.2 & 2.74 & 1.5 & 1.34 & 5718 & -1.12 & -0.72 &$ 3.5\cdot 10^{-5} $&$ 8.8\cdot 10^{-5} $& 4 \\
090902B & Fermi & 25 &  36.6 & 23.1 & 1.82 & 1.7 & 1.94 & 5762 & 1.09 &  0.55 &$ 1.4\cdot 10^{-4} $&$ 2.2\cdot 10^{-4} $&  - \\
100302A & Swift & 17.9 &  21.4 & 44.6 & 4.81 & 1.8 & 1.72 & 5675 & 0.04 &  0.03 &$ 1.4\cdot 10^{-3} $&$ 1.8\cdot 10^{-3} $&  - \\
100418A & Swift & 7.0 &  10.9 & 18.7 & 0.62 & 1.5 & 2.16 & 5978 & -1.33 & -0.77 &$ 2.9\cdot 10^{-5} $&$ 4.2\cdot 10^{-5} $&  - \\
110106B & Swift & 24.8 &  30.5 & 25.1 & 0.62 & 2.4 & 1.76 & 5675 & 2.25 &  2.09 &$ 2.5\cdot 10^{-4} $&$ 5.6\cdot 10^{-4} $&  - \\
110128A & Swift & 30.7 &  30.7 & 43.2 & 2.34 & 2.3 & 1.31 & 5675 & 2.39 &  2.39 &$ 1.1\cdot 10^{-3} $&$ 2.9\cdot 10^{-3} $& 16 \\
111211A & AGILE & 15 &  18.4 & 20.3 & 0.48 & 1.8 & 2.77 & 5545 & 0.78 &  0.92 &$ 1.8\cdot 10^{-4} $&$ 1.4\cdot 10^{-4} $&  - \\
120326A & Swift & 69.6 &  69.6 & 40.9 & 1.8 & 2.7 & CPL & 6021 & -0.8 & -0.80 &$  - $&$ 2.2\cdot 10^{-3} $&  - \\
120716A & IPN & 230 & 230.0 & 35.7 & 2.49 & 3.9 & CPL & 5718 & -0.57 & -0.57 &$  - $&$ 7.1\cdot 10^{-3} $&  - \\
120722A & Swift & 42.4 &  42.5 & 17.7 & 0.96 & 2.0 & 1.9 & 5848 & 1.23 &  1.17 &$ 1.8\cdot 10^{-4} $&$ 3.2\cdot 10^{-4} $&  - \\
120907A & Swift & 16.9 &  21.1 & 40.2 & 0.97 & 2.0 & 1.73 & 5892 & -1.55 & -1.49 &$ 2.4\cdot 10^{-4} $&$ 5.6\cdot 10^{-4} $&  - \\
130131B & Swift & 4.3 &  6.2 & 27.2 & 2.54 & 1.6 & 1.15 & 5762 & 0.85 &  0.50 &$ 5.2\cdot 10^{-5} $&$ 1.4\cdot 10^{-4} $& $<2$ \\
\enddata
\tablecomments{\\ $^a$Using the spectrum determined by satellites. \\
$^b$Assuming a differential spectral index 2.5. \\
$^c$99\% c.l.. \\
$^d$Derived from the $f_{sat}$ Fluence U.L. (see text). \\
$^e$For high energy emission extending up to 30 GeV only (see text).
\\
Column 1 is the GRB name corresponding to the detection date in
UT (YYMMDD). Column 2 gives the satellite that detected the burst. Column 3 gives the burst duration $\Delta t_{90}$ as measured
by the respective satellite. Column 4 gives the extended burst duration $\Delta t'_{90}$. Column 5 gives the zenith angle with respect to the detector location. Column 6 gives the GRB redshift. Column 7 gives the square root of the Fano factor. Column 8
reports the spectral index: "CPL" means that the spectrum measured by the satellite is better fitted with a cutoff power law. 
Column 9 gives the detector active area for that burst. Colums 10 and 11 give the statistical
significance of the on-source counts over the background for standard and extended burst duration. Columns 12 and 13 give the
99\% confidence upper limits on the fluence between 1 and 100 GeV for spectral index of Column 8 and fixed value -2.5, respectively. Column 14 gives the cutoff upper limit, if any.}
\label{tab:subb}
\end{deluxetable}

\begin{deluxetable}{clccccccrrccc}
\tabletypesize{\scriptsize}
\rotate
\tablecaption{GRBs with no measured redshift ($z$ = 2 and z=0.6 are assumed for long and short GRBs, respectively) observed by ARGO-YBJ.
}
\tablewidth{0pt}
\tablehead{
\colhead{GRB} & \colhead{Satellite} & 
\colhead{$\Delta t_{90}$} &
\colhead{$\Delta t'_{90}$} &
\colhead{$\theta $} & 
\colhead{$\sqrt{F}$} &
\colhead{Spectral Index} & 
\colhead{$A_{det}$} & \colhead{$\sigma$} & \colhead{$\sigma'$} &
\colhead{Fluence U.L.\tablenotemark{a,c}} &
\colhead{Fluence U.L.\tablenotemark{b,c}} &
\colhead{$E_{cut}$ U.L.\tablenotemark{d}} \\
\colhead{} & \colhead{} & \colhead{(s)} & \colhead{(s)} & \colhead{($^{\circ}$)} & \colhead{}  & \colhead{($\alpha_{sat}$)} & \colhead{(m$^2$)} & \colhead{}  & \colhead{}  &  \colhead{(erg cm$^{-2}$) ($f_{sat}$)} & \colhead{(erg cm$^{-2}$) ($f_{2.5}$)} & \colhead{(GeV)}
\\
\colhead{(1)} & \colhead{(2)} & \colhead{(3)} &
\colhead{(4)} & \colhead{(5)} & 
\colhead{(6)} & \colhead{(7)} & \colhead{(8)} &
\colhead{(9)} & \colhead{(10)} & \colhead{(11)} &
\colhead{(12)} & \colhead{(13)} 
}

\startdata
041228 & Swift & 55.4 & 68.4 & 28.1 & 1.5 & 1.6 & 563 & -0.01 & -0.27 &$ 5.8\cdot 10^{-4} $&$ 1.3\cdot 10^{-3} $& 52 \\
050509A & Swift & 11.4 &  16.9 & 34 & 1.3 & 2.11 & 1473 & 0.62 &  0.88 &$ 2.4\cdot 10^{-4} $&$ 3.5\cdot 10^{-4} $&  - \\
050528 & Swift & 11.3 &  17.3 & 37.8 & 1.2 & 2.27 & 1473 & 0.71 &  0.04 &$ 1.0\cdot 10^{-3} $&$ 1.2\cdot 10^{-3} $&  - \\
051105A & Swift & 0.1 &  - & 28.5 & 1.2 & 1.22 & 3119 & 1.24 &  - &$ 1.4\cdot 10^{-5} $&$ 4.5\cdot 10^{-5} $& 6 \\
051114 & Swift & 2 &  3.2 & 32.8 & 1.4 & 1.21 & 3032 & 3.37 & 3.27 &$ 5.0\cdot 10^{-5} $&$ 1.6\cdot 10^{-4} $& 5 \\
051227 & Swift & 114.6 & 114.6 & 22.8 & 2.9 & 1.45 & 2989 & 0.44 &  0.44 &$ 4.8\cdot 10^{-4} $&$ 1.1\cdot 10^{-3} $& 29 \\
060105 & Swift & 54.4 &  54.4 & 16.3 & 2.7 & 1.07 & 3119 & 1.77 &  1.77 &$ 3.3\cdot 10^{-4} $&$ 8.7\cdot 10^{-4} $& $<2$ \\
060111A & Swift & 13 &  19.6 & 10.8 & 1.5 & CPL & 3206 & 0.39 &  0.54 &$  - $&$ 1.1\cdot 10^{-4} $&  - \\
060121 & HETE & 2 & 3.3 & 41.9 & 1.3 & {\it 2.39} & 4159 & 0.6 & 0.58 &$ 2.2\cdot 10^{-4} $&$ 2.6\cdot 10^{-4} $&  - \\
060421 & Swift & 12 &  19.3 & 39.3 & 1.3 & 1.55 & 3855 & -0.51 & -0.62 &$ 2.7\cdot 10^{-4} $&$ 6.2\cdot 10^{-4} $& 31 \\
060424 & Swift & 37.5 &  46.5 & 6.7 & 1.7 & 1.71 & 4072 & 0.12 & -0.11 &$ 9.5\cdot 10^{-5} $&$ 1.8\cdot 10^{-4} $&  - \\
060427 & Swift & 64 &  76.6 & 32.6 & 1.8 & 1.87 & 4115 & -0.13 & -0.15 &$ 3.4\cdot 10^{-4} $&$ 6.0\cdot 10^{-4} $&  - \\
060510A & Swift & 20.4 &  29.9 & 37.4 & 1.6 & 1.57 & 3899 & 2.42 &  2.31 &$ 9.2\cdot 10^{-4} $&$ 2.1\cdot 10^{-3} $& 21 \\
060717 & Swift & 3 &  4.9 & 7.4 & 1.5 & 1.7 & 5155 & 1.58 &  0.33 &$ 2.9\cdot 10^{-5} $&$ 5.5\cdot 10^{-5} $&  - \\
060801 & Swift & 0.5 &  - & 16.8 & 1.3 & 0.47 & 5415 & 0.81 &  - &$ 7.4\cdot 10^{-6} $&$ 2.8\cdot 10^{-5} $& $<2$ \\
060805B & IPN & 8 &  12.1 & 29.1 & 1.5 & {\it 2.52} & 5285 & -0.45 & -0.17 &$ 1.3\cdot 10^{-4} $&$ 1.3\cdot 10^{-4} $&  - \\
060807 & Swift & 54 &  54.0 & 12.4 & 2.6 & 1.58 & 5155 & 0.78 &  0.78 &$ 1.7\cdot 10^{-4} $&$ 3.4\cdot 10^{-4} $& 43 \\
061028 & Swift & 106 & 106.0 & 42.5 & 1.9 & 1.73 & 5458 & -3.33 & -3.33 &$ 4.7\cdot 10^{-4} $&$ 9.5\cdot 10^{-4} $&  - \\
061122 & Integral & 18 &  18.0 & 33.5 & 3.9 & CPL & 5025 & 0.6 &  0.67 &$  - $&$ 7.4\cdot 10^{-4} $&  - \\
070201 & IPN & 0.3 &  - & 20.6 & 1.3 & CPL & 5242 & -1.21 &  - &$  - $&$ 1.1\cdot 10^{-5} $&  - \\
070219 & Swift & 16.6 &  20.0 & 39.3 & 1.8 & 1.78 & 4982 & -0.71 & -0.76 &$ 4.2\cdot 10^{-4} $&$ 8.0\cdot 10^{-4} $&  - \\
070306 & Swift & 209.5 & 209.5 & 19.9 & 3.4 & 1.66 & 2513 & -0.83 & -0.83 &$ 7.0\cdot 10^{-4} $&$ 1.3\cdot 10^{-3} $&  - \\
070531 & Swift & 44.5 &  58.0 & 44.3 & 1.6 & 1.41 & 2816 & 0.59 &  0.65 &$ 9.1\cdot 10^{-4} $&$ 2.3\cdot 10^{-3} $& 24 \\
070615 & Integral & 30 &  37.1 & 37.6 & 1.7 &  - & 5328 & 1.81 &  2.21 &$  - $&$ 2.0\cdot 10^{-3} $&  - \\
071013 & Swift & 26 &  32.1 & 13.3 & 1.9 & 1.6 & 4765 & -0.06 & -0.21 &$ 6.7\cdot 10^{-5} $&$ 1.4\cdot 10^{-4} $&  - \\
071101 & Swift & 9 &  14.1 & 32.8 & 1.4 & 2.25 & 3596 & 1.01 &  0.53 &$ 2.0\cdot 10^{-4} $&$ 2.5\cdot 10^{-4} $&  - \\
071104 & AGILE & 12 &  16.7 & 19.9 & 1.9 &  - & 4029 & -0.07 & -0.24 &$  - $&$ 1.5\cdot 10^{-4} $&  - \\
071118 & Swift & 71 &  71.0 & 41.2 & 2.8 & 1.63 & 5025 & 0.54 &  0.54 &$ 1.8\cdot 10^{-3} $&$ 3.9\cdot 10^{-3} $&  - \\
080328 & Swift & 90.6 &  90.6 & 37.2 & 2.7 & 1.52 & 6065 & -1.19 & -1.19 &$ 1.0\cdot 10^{-3} $&$ 2.4\cdot 10^{-3} $& 13 \\
080602 & Swift & 74 &  74.0 & 42 & 2.4 & 1.43 & 5762 & 1.24 &  1.24 &$ 1.5\cdot 10^{-3} $&$ 3.7\cdot 10^{-3} $& 18 \\
080613B & Swift & 105 & 105.0 & 39.2 & 2.6 & 1.39 & 5718 & 0.65 &  0.65 &$ 1.7\cdot 10^{-3} $&$ 4.3\cdot 10^{-3} $& 10 \\
080714B & Fermi & 5.4 & 8 & 24.4 & 1.5 & CPL & 5805 & -0.34 & -0.32 &$ - $&$ 5.8\cdot 10^{-5} $& - \\
080727C & Swift & 79.7 &  79.7 & 34.5 & 2.1 & CPL & 5415 & 3.52 &  3.52 &$  - $&$ 1.6\cdot 10^{-3} $&  - \\
080730A & Fermi & 17.4 & 25.5 & 31.2 & 1.5 & {\it 1.96} & 5545 & -0.26 & -0.87 &$ 1.2\cdot 10^{-4} $&$ 2.0\cdot 10^{-4} $& - \\
080822B & Swift & 64 &  65.8 & 40.3 & 2.4 & 2.54 & 5762 & -1.84 & -1.93 &$ 1.6\cdot 10^{-3} $&$ 1.5\cdot 10^{-3} $&  - \\
080830 & Fermi & 45 &  45.0 & 37.1 & 2.1 & {\it 1.69} & 5805 & -0.04 & -0.04 &$ 8.5\cdot 10^{-4} $&$ 1.8\cdot 10^{-3} $&  - \\
080903 & Swift & 66 &  66.0 & 21.5 & 2.3 & CPL & 5588 & -1.33 & -1.33 &$  - $&$ 2.6\cdot 10^{-4} $&  - \\
081025 & Swift & 23 &  32.6 & 30.5 & 1.6 & 1.12 & 5718 & -0.48 & -0.95 &$ 7.9\cdot 10^{-5} $&$ 2.3\cdot 10^{-4} $& $<2$ \\
081102B & Fermi & 2.2 &  3.3 & 27.8 & 1.4 & 1.07 & 5762 & 0.02 & -0.64 &$ 2.3\cdot 10^{-5} $&$ 6.8\cdot 10^{-5} $& $<2$ \\
081105 & IPN & 10 &  15.1 & 36.7 & 1.5 &  - & 5718 & -0.77 & -0.82 &$  - $&$ 4.7\cdot 10^{-4} $&  - \\
081122 & Fermi & 26 &  30.7 & 8.3 & 1.8 & {\it 2.24} & 4289 & -2.03 & -2.07 &$ 6.5\cdot 10^{-5} $&$ 8.1\cdot 10^{-5} $&  - \\
081128 & Swift & 100 & 100.0 & 31.8 & 3.6 & CPL & 5242 & -0.63 & -0.63 &$  - $&$ 1.1\cdot 10^{-3} $&  - \\
081130B & Fermi & 12 &  14.7 & 28.6 & 2.3 & CPL & 5978 & -0.05 &  0.03 &$  - $&$ 2.6\cdot 10^{-4} $&  - \\
081215A & Fermi & 7.7 &  10.3 & 35.9 & 1.9 & {\it 2.20} & 5762 & -0.15 &  0.26 &$ 4.5\cdot 10^{-4} $&$ 6.0\cdot 10^{-4} $&  - \\
090107A & Swift & 12.2 &  14.7 & 40.1 & 2.0 & 1.69 & 5762 & -1.12 & -1.59 &$ 3.0\cdot 10^{-4} $&$ 6.2\cdot 10^{-4} $&  - \\
090118 & Swift & 16 &  21.1 & 13.4 & 1.9 & 1.35 & 5805 & -1.62 & -1.45 &$ 2.7\cdot 10^{-5} $&$ 6.3\cdot 10^{-5} $& 3 \\
090301 & Swift & 41.0 &  41.0 & 14.2 & 2.5 & CPL & 5805 & 0.73 &  0.73 &$  - $&$ 2.6\cdot 10^{-4} $&  - \\
090301B & Fermi & 28 &  29.8 & 24.3 & 2.2 & {\it 1.93} & 5892 & -2.2 & -2.15 &$ 7.8\cdot 10^{-5} $&$ 1.2\cdot 10^{-4} $&  - \\
090306B & Swift & 20.4 &  20.4 & 38.5 & 2.3 & CPL & 5805 & -0.65 & -0.65 &$  - $&$ 1.1\cdot 10^{-3} $&  - \\
090320B & Fermi & 52 &  60.1 & 29 & 2.1 & CPL & 5892 & -0.25 &  0.04 &$  - $&$ 4.9\cdot 10^{-4} $&  - \\
090328B & Fermi & 0.32 &  - & 15.5 & 1.3 & {\it 2.48} & 5848 & 0.48 &  - &$ 1.6\cdot 10^{-5} $&$ 1.7\cdot 10^{-5} $&  - \\
090403 & Fermi & 16 &  21.5 & 28.5 & 1.8 &  - & 6021 & 0.65 &  1.16 &$  - $&$ 2.9\cdot 10^{-4} $&  - \\
090407 & Swift & 310 & 310.0 & 45 & 3.4 & 1.73 & 6021 & 1.53 &  1.53 &$ 6.7\cdot 10^{-3} $&$ 1.4\cdot 10^{-2} $&  - \\
090417B & Swift & 260 & 260.0 & 37.2 & 4.0 & 1.85 & 5978 & 0.64 &  0.64 &$ 6.2\cdot 10^{-3} $&$ 1.1\cdot 10^{-2} $&  - \\
090425 & Fermi & 72 &  92.0 & 44.6 & 1.9 & {\it 2.03} & 5848 & 1.7 &  2.13 &$ 2.1\cdot 10^{-3} $&$ 3.3\cdot 10^{-3} $&  - \\
090511 & Fermi & 14 &  17.9 & 39 & 1.7 & CPL & 5848 & 0.35 &  0.09 &$  - $&$ 8.8\cdot 10^{-4} $&  - \\
090520A & Swift & 20 &  25.1 & 42.2 & 2.0 & 0.8 & 5892 & -0.65 & -0.57 &$ 2.6\cdot 10^{-4} $&$ 9.2\cdot 10^{-4} $& $<2$ \\
090529C & Fermi & 10.4 &  15.6 & 22.1 & 1.4 & {\it 2.1} & 5892 & 1.16 &  1.34 &$ 8.8\cdot 10^{-5} $&$ 1.2\cdot 10^{-4} $&  - \\
090617 & Fermi & 0.45 &  - & 16.1 & 1.4 & {\it 2.00} & 5978 & 0.32 &  - &$ 1.4\cdot 10^{-5} $&$ 2.4\cdot 10^{-5} $&  - \\
090621B & Swift & 0.14 &  - & 40.5 & 1.3 & 0.82 & 5935 & 0.5 &  - &$ 2.4\cdot 10^{-5} $&$ 1.0\cdot 10^{-4} $& $<2$ \\
090704B & Fermi & 19.5 & 27.6 & 4.3 & 1.7 & 1.65 & 5848 & -0.66 & -0.37 &$ 3.7\cdot 10^{-5} $&$ 7.2\cdot 10^{-5} $& 28 \\
090712 & Swift & 145 & 145.0 & 10.7 & 3.7 & 1.33 & 5025 & -0.04 & -0.04 &$ 2.9\cdot 10^{-4} $&$ 6.9\cdot 10^{-4} $& 3 \\
090730A & Fermi & 9.1 & 14.6 & 4.4 & 1.4 & CPL & 5805 & 0.52 & -0.46 &$ - $&$ 5.9\cdot 10^{-5} $& - \\
090807A & Swift & 140.8 & 140.8 & 19.9 & 3.1 & 2.25 & 5935 & -0.76 & -0.76 &$ 5.1\cdot 10^{-4} $&$ 6.3\cdot 10^{-4} $&  - \\
090807B & Fermi & 3 &  5.2 & 29.3 & 1.3 & {\it 2.4} & 5978 & -1.14 & -2.69 &$ 5.1\cdot 10^{-5} $&$ 5.6\cdot 10^{-5} $&  - \\
090811A & Fermi & 14.8 & 21.7 & 23.1 & 1.7 & CPL & 5805 & -0.46 & -0.29 &$ - $&$ 1.1\cdot 10^{-4} $& - \\
090814B & Integral & 50 &  50.5 & 31.1 & 2.2 &  - & 5805 & -1.05 & -1.03 &$  - $&$ 4.2\cdot 10^{-4} $&  - \\
090817 & Integral & 220 & 220.0 & 14.6 & 2.9 & {\it 2.2} & 5892 & -0.77 & -0.77 &$ 4.0\cdot 10^{-4} $&$ 5.2\cdot 10^{-4} $&  - \\
090820A & Fermi & 30 &  41.2 & 17.1 & 1.7 & {\it 2.61} & 5935 & 0.25 &  0.39 &$ 2.0\cdot 10^{-4} $&$ 1.9\cdot 10^{-4} $&  - \\
090824A & Fermi & 59.9 & 69 & 30.8 & 1.7 & 2.01 & 5805 & 0.71 & 0.49 &$ 3.7\cdot 10^{-4} $&$ 5.8\cdot 10^{-4} $& - \\
090831A & Fermi & 53 &  67.9 & 35.8 & 2.8 & CPL & 4679 & 0.59 &  0.24 &$  - $&$ 3.4\cdot 10^{-3} $&  - \\
090904A & Swift & 122 & 164.8 & 21.9 & 1.9 & 2.01 & 5805 & 0.37 &  2.97 &$ 3.4\cdot 10^{-4} $&$ 5.1\cdot 10^{-4} $&  - \\
090904C & Fermi & 38.4 & 46.6 & 33 & 1.9 & CPL & 5978 & -1.66 & -1.87 &$ - $&$ 2.6\cdot 10^{-4} $& - \\
091106A & Fermi & 14.6 & 19.4 & 30.2 & 2 & CPL & 5762 & -0.17 & -0.32 &$ - $&$ 2.5\cdot 10^{-4} $& - \\
091202 & Integral & 45 &  45.0 & 33.2 & 3.2 &  - & 5415 & -0.64 & -0.64 &$  - $&$ 6.5\cdot 10^{-4} $&  - \\
091215A & Fermi & 4.4 & 6.1 & 25.4 & 1.5 & 1.65 & 5285 & -1.36 & -0.95 &$ 3.5\cdot 10^{-5} $&$ 7.4\cdot 10^{-5} $& 85 \\
091224A & Fermi & 0.8 & - & 16.8 &  1.3 & 1.21 & 5068 & -1.56 & - &$ 4.7\cdot 10^{-6} $&$ 1.4\cdot 10^{-5} $& $<2$ \\
091227A & Fermi & 21.9 & 23.6 & 27.9 & 2.1 & CPL & 5242 & 0.85 & 1.09 &$ - $&$ 4.4\cdot 10^{-4} $& - \\
100111A & Swift & 12.9 &  12.9 & 21.5 & 2.7 & 1.69 & 5458 & -1.03 & -1.03 &$ 7.5\cdot 10^{-5} $&$ 1.4\cdot 10^{-4} $&  - \\
100115A & Swift & 3 &  4.7 & 32.6 & 1.5 &  - & 5588 & -0.29 & -0.14 &$  - $&$ 8.4\cdot 10^{-5} $&  - \\
100122A & Fermi & 6.6 &  9.3 & 33.1 & 1.6 & {\it 2.31} & 5805 & 0.84 &  0.92 &$ 1.4\cdot 10^{-4} $&$ 1.7\cdot 10^{-4} $&  - \\
100131A & Fermi & 6.2 &  9.0 & 14 & 1.4 & {\it 2.21} & 5588 & 1.01 &  1.16 &$ 4.8\cdot 10^{-5} $&$ 6.1\cdot 10^{-5} $&  - \\
100206A & Swift & 0.12 &  - & 26.8 & 1.2 & 0.63 & 4245 & 0.9 &  - &$ 8.5\cdot 10^{-6} $&$ 3.4\cdot 10^{-5} $& $<2$ \\
100210A & Fermi & 29.2 & 29.2 & 24.9 & 2.4 & 1.71 & 5675 & 0.25 & 0.25 &$ 1.5\cdot 10^{-4} $&$ 2.7\cdot 10^{-4} $& - \\
100225B & Fermi & 32.0 & 38.4 & 22.1 & 1.9 & 1.51 & 5892 & -1.83 & -2.24 &$ 5.9\cdot 10^{-5} $&$ 1.2\cdot 10^{-4} $& 28 \\
100225D & Fermi & 4.5 & 7.1 & 8.4 & 1.5 & CPL & 5805 & -1.19 & -1.11 &$ - $&$ 2.8\cdot 10^{-5} $& - \\
100424A & Swift & 104 & 104.0 & 33.4 & 2.4 & 1.83 & 6021 & 0.41 &  0.41 &$ 5.3\cdot 10^{-4} $&$ 9.6\cdot 10^{-4} $&  - \\
100503A & Fermi & 129.5 & 129.5 & 26.4 & 4.5 & CPL & 6065 & 0.09 &  0.09 &$  - $&$ 1.9\cdot 10^{-3} $&  - \\
100513B & Fermi & 11.1 & 16.2 & 38.7 & 1.6 & CPL & 5502 & 1.16 & 1.17 &$ - $&$ 9.0\cdot 10^{-4} $& - \\
100522A & Swift & 35.3 &  46.0 & 27.7 & 1.7 & 1.89 & 4679 & 0.86 &  0.62 &$ 2.8\cdot 10^{-4} $&$ 4.9\cdot 10^{-4} $&  - \\
100525A & Fermi & 1.5 & 2.5 & 13.7 & 1.4 & CPL & 5892 & 0.65 & -0.16 &$ - $&$ 2.2\cdot 10^{-5} $& - \\
100526A & Swift & 102 & 102.7 & 9.5 & 2.5 & 1.83 & 5935 & -0.72 & -0.79 &$ 1.7\cdot 10^{-4} $&$ 2.8\cdot 10^{-4} $&  - \\
100527A & Fermi & 184.6 & 297.4 & 33.3 & 2.2 & CPL & 5935 & 2.33 & 3.4 &$ - $&$ 2.6\cdot 10^{-3} $& - \\
100530A & Fermi & 3.3 & 5 & 39 & 1.4 & 1.66 & 6108 & 1.15 & 0.14 &$ 1.8\cdot 10^{-4} $&$ 3.9\cdot 10^{-4} $& - \\
100614B & Fermi & 172.3 & 197.9 & 43.2 & 2.4 & CPL & 5978 & -0.24 & -0.86 &$ - $&$ 4.0\cdot 10^{-3} $& - \\
100621C & Fermi & 1.0 & - & 31.5 &  1.3 & - & 5762 & 0.71 & - &$ - $&$ 4.3\cdot 10^{-5} $& - \\
100625B & Fermi & 29.2 & 36.4 & 15.4 & 1.7 & CPL & 5458 & 1.07 & 0.55 &$ - $&$ 2.5\cdot 10^{-4} $& - \\
100706A & Fermi & 0.1 & - & 18.3 &  1.3 & 1.28 & 5675 & 0.39 & - &$ 5.6\cdot 10^{-6} $&$ 1.6\cdot 10^{-5} $& 4 \\
100713A & Integral & 20 &  23.1 & 12.5 & 2.6 &  - & 5848 & 0.43 &  0.37 &$  - $&$ 1.7\cdot 10^{-4} $&  - \\
100714B & Fermi & 5.6 & 8.7 & 39.9 & 1.5 & CPL & 5502 & -0.81 & -0.07 &$ - $&$ 3.6\cdot 10^{-4} $& - \\
100718A & Fermi & 38.7 & 47.6 & 16.7 & 2.1 & CPL & 5632 & -0.52 & -0.47 &$ - $&$ 2.2\cdot 10^{-4} $& - \\
100728A & Swift & 198.5 & 198.5 & 44.8 & 2.6 & 1.18 & 6021 & 0.49 &  0.49 &$ 2.0\cdot 10^{-3} $&$ 5.8\cdot 10^{-3} $& $<2$ \\
100902A & Swift & 428.8 & 428.8 & 37 & 5.0 & 1.98 & 5415 & 0.41 &  0.41 &$ 1.2\cdot 10^{-2} $&$ 1.9\cdot 10^{-2} $&  - \\
100929A & Fermi & 8.2 & 12.4 & 34.9 & 1.4 & 1.36 & 5892 & -0.38 & -0.68 &$ 4.8\cdot 10^{-5} $&$ 1.2\cdot 10^{-4} $& 9 \\
100929B & Fermi & 4.6 & 7.2 & 27.2 & 1.5 & 1.54 & 5848 & -0.68 & -0.61 &$ 4.0\cdot 10^{-5} $&$ 9.1\cdot 10^{-5} $& - \\
101003A & Fermi & 10.0 & 15.3 & 30.8 & 1.5 & CPL & 5892 & -0.28 & -0.11 &$ - $&$ 1.5\cdot 10^{-4} $& - \\
101008A & Swift & 104 & 104.0 & 25.6 & 2.4 & 1.59 & 5848 & 0.3 &  0.30 &$ 4.5\cdot 10^{-4} $&$ 9.8\cdot 10^{-4} $&  - \\
101101A & Fermi & 3.3 & 4.8 & 25.5 & 1.5 & 2.02 & 5935 & -0.02 & 0.57 &$ 5.5\cdot 10^{-5} $&$ 8.6\cdot 10^{-5} $& - \\
101107A & Fermi & 375.8 & 375.8 & 25.8 & 3 & CPL & 5892 & 3 & 3 &$ - $&$ 6.4\cdot 10^{-3} $& - \\
101112B & Fermi & 82.9 & 82.9 & 39.9 & 3.3 & CPL & 5805 & -0.95 & -0.95 &$ - $&$ 3.0\cdot 10^{-3} $& - \\
101123A & Fermi & 105 & 105.0 & 23.7 & 3.5 & {\it 2.14} & 5978 & -1.79 & -1.79 &$ 3.1\cdot 10^{-4} $&$ 4.1\cdot 10^{-4} $&  - \\
101202A & Fermi & 18.4 & 26.3 & 38 & 1.5 & 1.62 & 5848 & -0.61 & -0.38 &$ 3.0\cdot 10^{-4} $&$ 6.6\cdot 10^{-4} $& - \\
101208A & Fermi & 0.2 & - & 37.3 &  1.2 & CPL & 3899 & -0.42 & - &$ - $&$ 8.9\cdot 10^{-5} $& - \\
101224A & Swift & 0.2 &  - & 22.6 & 1.3 & CPL & 5675 & -0.67 &  - &$  - $&$ 1.2\cdot 10^{-5} $&  - \\
101231A & Fermi & 23.6 & 23.6 & 24 & 2.5 & {\it 2.44} & 5675 & -0.58 & -0.58 &$ 1.9\cdot 10^{-4} $&$ 1.9\cdot 10^{-4} $& - \\
110101A & Fermi & 3.6 & 5 & 6.4 & 1.8 & 1.51 & 5848 & 0.21 & 0.23 &$ 2.1\cdot 10^{-5} $&$ 4.4\cdot 10^{-5} $& - \\
110106A & Swift & 4.3 &  6.0 & 34.8 & 1.6 & 1.71 & 5588 & -1.28 & -1.71 &$ 3.8\cdot 10^{-5} $&$ 7.6\cdot 10^{-5} $&  - \\
110206B & Fermi & 12.3 & 17.8 & 43.4 & 1.5 & 1.55 & 5458 & -0.07 & 0.13 &$ 2.7\cdot 10^{-4} $&$ 6.2\cdot 10^{-4} $& - \\
110210A & Swift & 233 & 233.0 & 23 & 7.2 & 1.73 & 5762 & 1.15 &  1.15 &$ 1.9\cdot 10^{-3} $&$ 3.4\cdot 10^{-3} $&  - \\
110220A & Fermi & 33.0 & 33 & 31 & 2.2 & CPL & 5935 & 1.83 & 1.83 &$ - $&$ 7.4\cdot 10^{-4} $& - \\
110226A & Fermi & 14.1 & 17.7 & 37 & 1.8 & CPL & 5805 & -0.89 & -1.09 &$ - $&$ 6.4\cdot 10^{-4} $& - \\
110312A & Swift & 28.7 &  30.3 & 37.2 & 2.2 & 2.32 & 5805 & 0.3 &  0.21 &$ 1.3\cdot 10^{-3} $&$ 1.5\cdot 10^{-3} $&  - \\
110315A & Swift & 77 &  85.9 & 19.3 & 2.9 & 1.77 & 5112 & -2.26 & -2.58 &$ 1.6\cdot 10^{-4} $&$ 2.9\cdot 10^{-4} $&  - \\
110328B & Fermi & 40 &  40.0 & 20.8 & 2.6 & 3.31 & 6151 & 1.34 &  1.34 &$ 7.4\cdot 10^{-4} $&$ 4.4\cdot 10^{-4} $&  - \\
110401A & Fermi & 2 & 3.5 & 15.2 & 1.3 & {\it 2.36} & 5675 & -0.94 & -0.35 &$ 1.9\cdot 10^{-5} $&$ 2.2\cdot 10^{-5} $&  - \\
110406A & Integral & 8 &  12.7 & 31.1 & 1.5 & {\it 2.30} & 6108 & -0.1 &  0.07 &$ 1.1\cdot 10^{-4} $&$ 1.3\cdot 10^{-4} $&  - \\
110414A & Swift & 152.0 & 152.0 & 44.1 & 3.3 & 1.7 & 6021 & -0.92 & -0.92 &$ 2.0\cdot 10^{-3} $&$ 4.2\cdot 10^{-3} $&  - \\
110517A & Fermi & 0.6 & - & 29.5 &  1.3 & 1.29 & 5198 & 2.55 & - &$ 2.1\cdot 10^{-5} $&$ 6.8\cdot 10^{-5} $& 13 \\
110605A & Fermi & 82.7 & 82.7 & 33.8 & 3 & {\it 2.20} & 6065 & -0.01 & -0.01 &$ 7.0\cdot 10^{-4} $&$ 9.3\cdot 10^{-4} $& - \\
110605B & Fermi & 1.5 & 2.6 & 39.9 & 1.4 & 1.5 & 5935 & 0.21 & 0.5 &$ 5.8\cdot 10^{-5} $&$ 1.7\cdot 10^{-4} $& 46 \\
110625A & Swift & 44.5 &  52.3 & 40 & 2.0 & 1.44 & 5892 & -1.15 & -1.38 &$ 4.9\cdot 10^{-4} $&$ 1.2\cdot 10^{-3} $& 3 \\
110626A & Fermi & 6.4 & 9.9 & 40.4 & 1.4 & CPL & 5892 & 0.19 & 0.41 &$ - $&$ 4.3\cdot 10^{-4} $& - \\
110629A & Fermi & 61.7 & 70.6 & 5.1 & 1.9 & CPL & 6065 & 1.23 & 1.53 &$ - $&$ 2.8\cdot 10^{-4} $& - \\
110705B & Fermi & 19.2 & 29.2 & 18.8 & 1.6 & CPL & 5892 & -1.09 & -1.28 &$ - $&$ 9.5\cdot 10^{-5} $& - \\
110709A & Swift & 44.7 &  46.7 & 13.5 & 2.3 & 1.24 & 6021 & 0.48 &  0.34 &$ 9.4\cdot 10^{-5} $&$ 2.3\cdot 10^{-4} $& $<2$ \\
110709C & Fermi & 24.1 & 32.6 & 26.7 & 2 & CPL & 5935 & -1.39 & -1.21 &$ - $&$ 2.2\cdot 10^{-4} $& - \\
110820A & Swift & 256 & 256.0 & 41.6 & 4.7 & 1.92 & 5978 & 1.63 &  1.63 &$ 9.6\cdot 10^{-3} $&$ 1.6\cdot 10^{-2} $&  - \\
110915A & Swift & 78.8 &  95.7 & 39.5 & 1.9 & CPL & 5848 & 1.01 &  0.65 &$  - $&$ 2.8\cdot 10^{-3} $&  - \\
110919A & Fermi & 35.1 & 46.3 & 42.6 & 1.7 & CPL & 5848 & 0.04 & 0.33 &$ - $&$ 1.3\cdot 10^{-3} $& - \\
110921A & Swift & 48.0 &  55.6 & 7.2 & 2.1 & 1.57 & 5762 & 1.98 &  1.82 &$ 1.6\cdot 10^{-4} $&$ 3.3\cdot 10^{-4} $& 12 \\
110928B & Fermi & 148.2 & 161 & 8.5 & 2 & {\it 1.92} & 5068 & 0.26 & -0.03 &$ 2.5\cdot 10^{-4} $&$ 4.0\cdot 10^{-4} $& - \\
111017A & Fermi & 11.1 & 17.7 & 40 & 1.4 & CPL & 5892 & -2 & -1.16 &$ - $&$ 3.3\cdot 10^{-4} $& - \\
111024C & Fermi & 1.8 & 2.6 & 32.2 & 1.2 & CPL & 3812 & -0.99 & 0.26 &$ - $&$ 3.7\cdot 10^{-5} $& - \\
111103B & Swift & 167 & 167.0 & 41.6 & 3.0 & 1.41 & 5892 & 1.6 &  1.60 &$ 3.2\cdot 10^{-3} $&$ 8.1\cdot 10^{-3} $& 14 \\
111109C & Fermi & 9.7 & 11.9 & 32 & 1.9 & CPL & 5848 & 0.79 & 0.8 &$ - $&$ 2.5\cdot 10^{-4} $& - \\
111113A & IPN & 0.5 &  - & 28.4 & 1.4 & CPL & 5805 & 0.26 &  - &$  - $&$ 3.8\cdot 10^{-5} $&  - \\
111208A & Swift & 20 &  20.2 & 11.1 & 2.6 & 1.5 & 5112 & -0.97 & -0.97 &$ 5.6\cdot 10^{-5} $&$ 1.2\cdot 10^{-4} $& 6 \\
111215A & Swift & 796 & 796.0 & 30.6 & 23.5 & 1.7 & 5848 & 0.65 &  0.65 &$ 2.0\cdot 10^{-2} $&$ 4.0\cdot 10^{-2} $&  - \\
111228B & Fermi & 2.9 & 4.2 & 23.9 & 1.4 & CPL & 5588 & -1.26 & 0.13 &$ - $&$ 2.9\cdot 10^{-5} $& - \\
120102A & Swift & 38.7 &  38.7 & 44.8 & 2.4 & 1.59 & 5545 & 1.51 &  1.51 &$ 1.3\cdot 10^{-3} $&$ 2.8\cdot 10^{-3} $&  - \\
120106A & Swift & 61.6 &  61.6 & 35.4 & 2.6 & 1.53 & 5588 & -0.24 & -0.24 &$ 1.0\cdot 10^{-3} $&$ 2.4\cdot 10^{-3} $&  - \\
120118B & Swift & 23.3 &  23.3 & 42.7 & 2.5 & 2.08 & 5502 & 0.79 &  0.79 &$ 1.3\cdot 10^{-3} $&$ 1.9\cdot 10^{-3} $&  - \\
120118C & Fermi & 17.2 & 17.2 & 18.1 & 2.3 & CPL & 5458 & 1.1 & 1.1 &$ - $&$ 2.5\cdot 10^{-4} $& - \\
120129A & IPN & 4 &  6.3 & 38.5 & 1.5 & {\it 2.9} & 5718 & -0.07 &  0.29 &$ 5.0\cdot 10^{-4} $&$ 3.5\cdot 10^{-4} $&  - \\
120202A & Integral & 100 & 104.1 & 15.6 & 4.0 &  - & 5718 & -0.14 & -0.23 &$  - $&$ 7.6\cdot 10^{-4} $&  - \\
120217A & Fermi & 5.9 & 8.4 & 38.8 & 1.5 & CPL & 5458 & 0.79 & 1.03 &$ - $&$ 5.3\cdot 10^{-4} $& - \\
120219A & Swift & 90.5 &  90.5 & 32 & 3.4 & CPL & 5545 & -0.56 & -0.56 &$  - $&$ 9.8\cdot 10^{-4} $&  - \\
120222A & Fermi & 1.1 & - & 44 &  1.4 & CPL & 5588 & 0.43 & - &$ - $&$ 1.5\cdot 10^{-4} $& - \\
120223A & Fermi & 14.3 & 16 & 37.6 & 2 & CPL & 5632 & 0.39 & 0.49 &$ - $&$ 1.0\cdot 10^{-3} $& - \\
120226B & Fermi & 14.6 & 18.1 & 36.8 & 1.9 & CPL & 5632 & -0.64 & -0.77 &$ - $&$ 7.6\cdot 10^{-4} $& - \\
120509A & Fermi & 0.7 & - & 14.2 &  1.3 & - & 5892 & -0.33 & - &$ - $&$ 1.3\cdot 10^{-5} $& - \\
120512A & Integral & 40 &  47.2 & 36.8 & 1.9 & CPL & 5892 & 0.06 & -0.29 &$  - $&$ 1.5\cdot 10^{-3} $&  - \\
120519A & IPN & 1.2 &  - & 44.8 & 1.3 & CPL & 5935 & -2.02 &  - &$  - $&$ 6.9\cdot 10^{-5} $&  - \\
120522B & Fermi & 28.2 & 38.9 & 40.2 & 1.6 & {\it 2.04} & 5892 & 0.75 & 1.16 &$ 8.4\cdot 10^{-4} $&$ 1.3\cdot 10^{-3} $& - \\
120604B & Fermi & 12.0 & 16.9 & 33.5 & 1.6 & 1.73 & 5892 & 0.91 & 0.63 &$ 1.2\cdot 10^{-4} $&$ 2.4\cdot 10^{-4} $& - \\
120612B & Fermi & 63.2 & 71.4 & 21.5 & 1.8 & 1.57 & 4375 & 0.68 & 0.83 &$ 2.0\cdot 10^{-4} $&$ 4.0\cdot 10^{-4} $& - \\
120625A & Fermi & 7.4 & 11.1 & 21.2 & 1.5 & {\it 2.30} & 5805 & -0.27 & -0.46 &$ 5.8\cdot 10^{-5} $&$ 6.8\cdot 10^{-5} $& - \\
120630A & Swift & 0.6 &  - & 13.6 & 1.3 & 1.04 & 5632 & -0.78 &  - &$ 3.5\cdot 10^{-6} $&$ 1.1\cdot 10^{-5} $& $<2$ \\
120703C & Fermi & 77.6 & 77.6 & 21.8 & 2.5 & 1.68 & 5675 & 0.32 & 0.32 &$ 2.6\cdot 10^{-4} $&$ 4.9\cdot 10^{-4} $& - \\
120727354 & Fermi & 0.90 & - & 5 &  1.3 & - & 5892 & -1.25 & - &$ - $&$ 9.8\cdot 10^{-6} $& - \\
120819A & Swift & 71 &  71.0 & 42.1 & 2.8 & 1.49 & 5892 & 0.55 &  0.55 &$ 1.5\cdot 10^{-3} $&$ 3.6\cdot 10^{-3} $&  - \\
120905657 & Fermi & 195.6 & 235.2 & 41.8 & 4.4 & - & 5935 & -2.03 & -1.66 &$ - $&$ 4.9\cdot 10^{-3} $& - \\
120915474 & Fermi & 5.9 & 8.3 & 40.9 & 1.5 & - & 5935 & -0.08 & -0.2 &$ - $&$ 4.2\cdot 10^{-4} $& - \\
121011A & Swift & 75.6 &  75.6 & 19.3 & 2.2 & CPL & 5805 & -0.1 & -0.10 &$  - $&$ 3.6\cdot 10^{-4} $&  - \\
121012A & Fermi & 0.45 &  - & 24.7 & 1.3 & CPL & 5805 & 0.7 &  - &$  - $&$ 2.4\cdot 10^{-5} $&  - \\
121025A & MAXI/ISS & 20 &  30.6 & 6.9 & 1.4 &  - & 3596 & 0.52 &  0.38 &$  - $&$ 1.2\cdot 10^{-4} $&  - \\
121108A & Swift & 89 &  89.0 & 36.1 & 3.3 & 2.28 & 5718 & -0.19 & -0.19 &$ 3.1\cdot 10^{-3} $&$ 3.8\cdot 10^{-3} $&  - \\
121113544 & Fermi & 95.5 & 95.5 & 34.5 & 3.1 & - & 5718 & 0.6 & 0.6 &$ - $&$ 1.3\cdot 10^{-3} $& - \\
121123A & Swift & 317 & 317.0 & 42.1 & 4.9 & CPL & 5935 & -0.25 & -0.25 &$  - $&$ 1.2\cdot 10^{-2} $&  - \\
121202A & Swift & 20.1 &  24.1 & 27.2 & 2.1 & 1.59 & 5632 & 0.67 &  0.84 &$ 1.8\cdot 10^{-4} $&$ 3.9\cdot 10^{-4} $& 34 \\
130116415 & Fermi & 66.8 & 66.8 & 41.1 & 3.4 & - & 5718 & -0.24 & -0.24 &$ - $&$ 3.3\cdot 10^{-3} $& - \\
130122A & Swift & 64 &  64.0 & 30.6 & 2.4 & 1.34 & 5762 & -0.23 & -0.23 &$ 2.3\cdot 10^{-4} $&$ 6.0\cdot 10^{-4} $& 7 \\
\enddata
\tablecomments{\\$^a$Using the spectrum determined by satellites. \\
$^b$Assuming a differential spectral index 2.5. \\
$^c$99\% c.l.. \\
$^d$Derived from the $f_{sat}$ Fluence U.L. (see text).
\\
Column 1 is the GRB name corresponding to the detection date in
UT (YYMMDD). Column 2 gives the satellite that detected the burst. Column 3 gives the burst duration $\Delta t_{90}$ as measured
by the respective satellite. Column 4 gives the extended burst duration $\Delta t'_{90}$. Column 5 gives the zenith angle with respect to the detector location. Column 6 gives the square root of the Fano factor. Column 7
reports the spectral index: "CPL" means that the spectrum measured by the satellite is better fitted with a cutoff power law. In case
of double power law fit (Band or SBPL functions) the higher energy spectral 
index is reported (in italics). 
Column 8 gives the detector active area for that burst. Colums 9 and 10 give the statistical
significance of the on-source counts over the background for standard and extended burst duration. Columns 11 and 12 give the
99\% confidence upper limits on the fluence between 1 and 100 GeV for spectral index of Column 7 and fixed value -2.5, respectively. Column 13 gives the cutoff upper limit, if any.}
\label{tab:all}
\end{deluxetable}

\end{document}